\PassOptionsToPackage{table}{xcolor}
\documentclass[acmsmall]{acmart}

\usepackage{graphicx}   
\usepackage{booktabs}   
\usepackage{array}      
\usepackage{tabularx}   
\usepackage{longtable}  
\usepackage{colortbl}   
\usepackage{xcolor}     
\usepackage{makecell}   
\usepackage{float}      
\usepackage{ragged2e}   
\usepackage{pdflscape}  

\definecolor{tabheader}{RGB}{70,172,200} 

\newcolumntype{P}[1]{>{\RaggedRight\arraybackslash}p{#1}} 
\newcolumntype{L}[1]{>{\RaggedRight\arraybackslash}p{#1}} 

\DeclareUnicodeCharacter{202F}{\,} 

\AtBeginDocument{%
  }

\setcopyright{acmlicensed}
\copyrightyear{2025}
\acmYear{2025}
\acmDOI{XXXXXXX.XXXXXXX}

\acmJournal{CSUR}

\begin{document}

\title{Post-Quantum Cryptography and Quantum-Safe Security: A Comprehensive Survey}


\author{Gaurab Chhetri}
\email{gaurab@txstate.edu}
\orcid{0009-0000-0124-4814}
\affiliation{%
  \institution{Texas State University}
  \city{San Marcos}
  \state{Texas}
  \country{USA}
}

\author{Shriyank Somvanshi}
\affiliation{%
  \institution{Texas State University}
  \city{San Marcos}
  \country{USA}}
\email{shriyank@txstate.edu}

\author{Pavan Hebli}
\affiliation{%
  \institution{Texas State University}
  \city{San Marcos}
  \country{USA}}
\email{zea16@txstate.edu}

\author{Shamyo Brotee}
\affiliation{%
  \institution{Texas State University}
  \city{San Marcos}
  \country{USA}}
\email{s.brotee@txstate.edu}

\author{Subasish Das, Ph.D.}
\affiliation{%
  \institution{Texas State University}
  \city{San Marcos}
  \country{USA}}
\email{subasish@txstate.edu}

\renewcommand{\shortauthors}{Chhetri et al.}

\begin{abstract}
Post-quantum cryptography (PQC) is moving from evaluation to deployment as NIST finalizes standards for ML-KEM, ML-DSA, and SLH-DSA. This survey maps the space from foundations to practice. We first develop a taxonomy across lattice-, code-, hash-, multivariate-, isogeny-, and MPC-in-the-Head families, summarizing security assumptions, cryptanalysis, and standardization status. We then compare performance and communication costs using representative, implementation-grounded measurements, and review hardware acceleration (AVX2, FPGA/ASIC) and implementation security with a focus on side-channel resistance. Building upward, we examine protocol integration (TLS, DNSSEC), PKI and certificate hygiene, and deployment in constrained and high-assurance environments (IoT, cloud, finance, blockchain). We also discuss complementarity with quantum technologies (QKD, QRNGs) and the limits of near-term quantum computing. Throughout, we emphasize crypto-agility, hybrid migration, and evidence-based guidance for operators. We conclude with open problems spanning parameter agility, leakage-resilient implementations, and domain-specific rollout playbooks. This survey aims to be a practical reference for researchers and practitioners planning quantum-safe systems, bridging standards, engineering, and operations.
\end{abstract}



\received{11 October 2025}

\maketitle

\section{Introduction}
\label{sec:introduction}

Post-quantum cryptography (PQC) has become the primary defense against the vulnerabilities that large-scale quantum computing introduces to traditional cryptographic systems. Classical public-key schemes such as Rivest–Shamir–Adleman (RSA) and elliptic curve cryptography (ECC) are expected to be broken once quantum adversaries are realized, which makes proactive cryptographic migration a necessity. While practical quantum computers are still under development, governments, standardization agencies, and industry stakeholders are already working on transition strategies to safeguard sensitive information and mission-critical infrastructure for the long term~\cite{mamatha2024post}.

The sense of urgency surrounding PQC adoption is amplified by the “harvest now, decrypt later” (HNDL) scenario, where adversaries may intercept and store encrypted data today with the intent of decrypting it once quantum capabilities mature~\cite{joseph2022transitioning}. This paradigm shift forces policymakers and researchers to rethink the very foundation of digital trust and to anticipate a future in which long-term confidentiality must be guaranteed even against future computational advances. Critical sectors such as defense, finance, and healthcare are particularly vulnerable, as the exposure of archived data could have irreversible national security and privacy implications. To mitigate this looming threat, the U.S. National Institute of Standards and Technology (NIST) launched its post-quantum cryptography standardization initiative in 2016, marking one of the most extensive collaborative efforts in modern cryptographic history. Over several rounds of public evaluation, peer review, and international collaboration, NIST rigorously assessed algorithmic performance, cryptanalytic resistance, and implementation efficiency. The process culminated in the formal approval of three algorithms (e.g., CRYSTALS-Kyber, CRYSTALS-Dilithium, and SPHINCS+) as Federal Information Processing Standards (FIPS) in 2024~\cite{nist_fips_pqc_2024,alagic2025status}. These algorithms collectively represent the culmination of years of research in lattice- and hash-based cryptography, offering strong resilience against both classical and quantum adversaries.

The adoption of these standards underscores a significant milestone in the evolution of modern cryptography. Lattice-based schemes, exemplified by Kyber and Dilithium, provide a balance of computational efficiency and security grounded in the hardness of structured lattice problems, while SPHINCS+ highlights the versatility of hash-based designs for digital signatures. Despite these advances, other promising cryptographic families—including multivariate and isogeny-based approaches—remain subjects of active exploration, refinement, and in some cases, cryptanalytic challenges~\cite{beullens2022breaking,castryck2023efficient}. These developments illustrate that PQC is not a single algorithmic solution but a dynamic research frontier where mathematical innovation, hardware optimization, and implementation security converge.

In this evolving landscape, the global cryptographic community faces a dual challenge: ensuring that PQC algorithms are not only theoretically sound but also practical for real-world deployment across diverse systems and protocols. As nations and industries prepare for a post-quantum era, the focus is shifting from algorithm selection to ecosystem integration, implementation security, and lifecycle management. This paper contributes to this growing body of knowledge by examining the emerging design patterns, evaluation frameworks, and transition pathways that define the practical realization of quantum-safe cryptography. 

\subsection{Scope of the Survey}

This paper surveys PQC from both algorithmic and system perspectives. At the algorithmic level, it provides a taxonomy of candidate families, comparing their theoretical underpinnings, efficiency characteristics, and security assumptions~\cite{hosseini2024comprehensive,benny_meta_analysis_pqc_2024}. At the system level, it reviews deployment considerations including hardware performance, crypto-agility, and domain-specific challenges in financial services, internet of things (IoT) ecosystems, and blockchain platforms~\cite{liu2024post,gsma2025PQIoT,pape2024beyondPQC,allende2023quantum}. Hybrid approaches, which combine classical and post-quantum primitives, are also discussed as transitional strategies for maintaining interoperability and reducing adoption risks~\cite{schwabe2021kemtls,zacharopoulos2024drawbacks}.  

\subsection{Summary of Contributions}

The contributions of this survey are fourfold. First, we classify and analyze the major PQC algorithmic families, highlighting their mathematical foundations, strengths, weaknesses, and progress through the NIST standardization process.
Second, we examine system-level migration challenges, including performance optimization, side-channel resistance, and integration into resource-constrained or high-assurance environments~\cite{demir2025performance,commey2025performance}.
Third, we synthesize open research directions, particularly in hybrid deployments, crypto-agility frameworks, and the interaction of PQC with complementary technologies such as Quantum Key Distribution (QKD)~\cite{garg2024post,garms2024experimental}.
Finally, to promote accessibility and continuous learning, we introduce \textit{awesome-pqc}\footnote{Awesome PQC repository: \url{https://github.com/gauravfs-14/awesome-pqc}}, a curated open repository that compiles state-of-the-art resources, research papers, libraries, and tools for the post-quantum cryptography community. This living repository serves as a practical companion to this survey and a long-term reference point for researchers and practitioners navigating the evolving landscape of quantum-safe cryptography.

The remainder of this paper is structured as follows. Section~\ref{sec:taxonomy} introduces the taxonomy and foundational principles of post-quantum cryptography (PQC). Section~\ref{sec:algo-fam} outlines the core algorithmic families that form the basis of PQC, while Section~\ref{sec:digital-schemes} focuses specifically on digital signature schemes. Section~\ref{sec:perf-eval} provides a detailed performance evaluation across representative algorithms. Section~\ref{sec:sys-lvl-considerations} discusses system-level considerations for PQC integration, followed by Section~\ref{sec:domain-spec}, which explores domain-specific implications in diverse application contexts. Section~\ref{sec:quantum-sec} examines the complementary role of quantum technologies in enhancing security frameworks. Section~\ref{sec:comp-analysis} presents a comparative discussion consolidating key insights from the preceding sections. Finally, Section~\ref{sec:open-problems} highlights open challenges and directions for future research, and Section~\ref{sec:conclusion} concludes the paper. The acronyms used throughout are summarized in~\autoref{tab:acronym} (see Appendix).

\section{Taxonomy and Foundations}
\label{sec:taxonomy}

\begin{figure}
    \centering
    \includegraphics[width=0.96\linewidth]{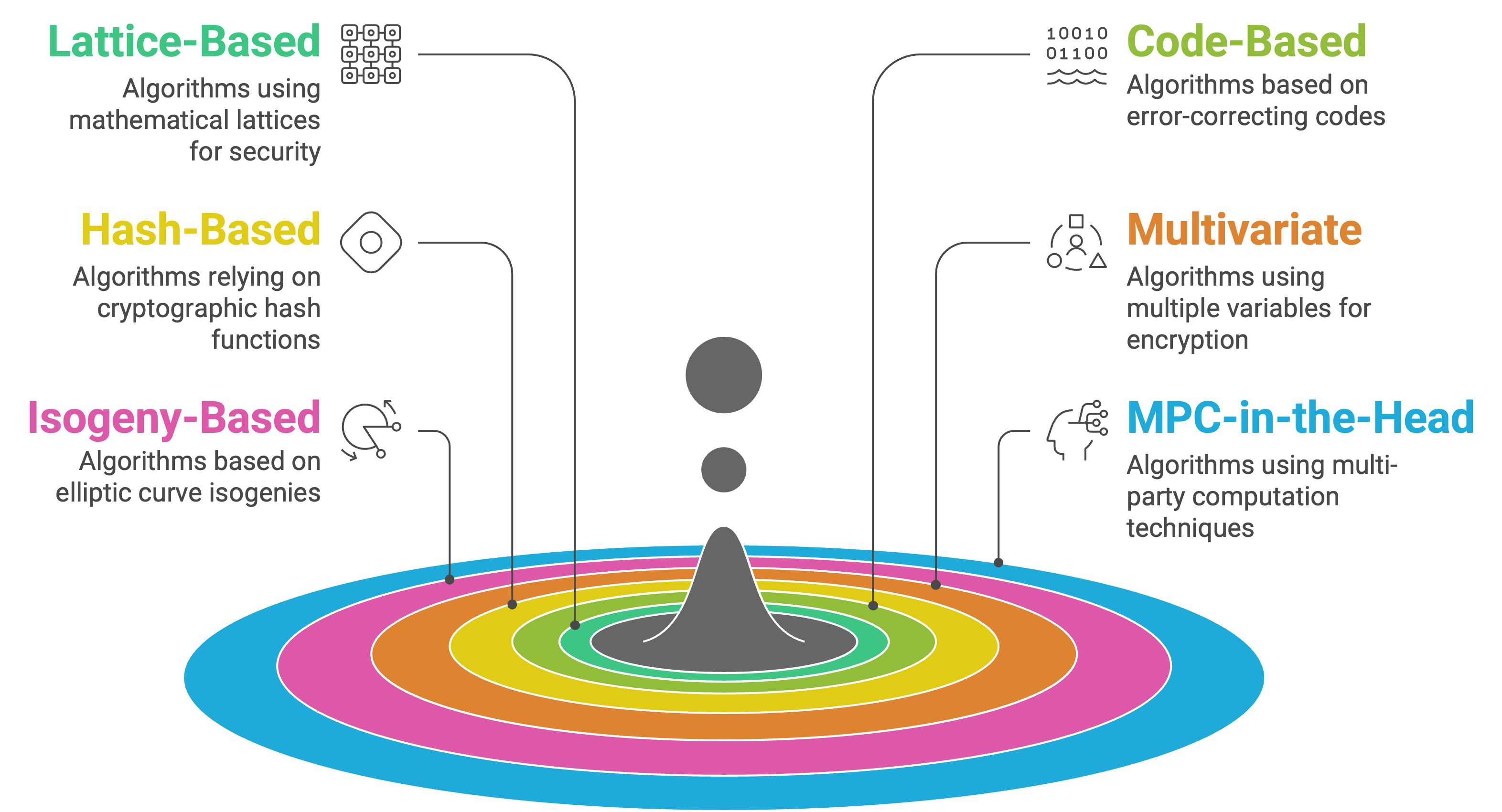}
    \caption{Taxonomy of major PQC algorithm families. The classification highlights six primary categories-lattice-based, code-based, hash-based, multivariate, isogeny-based, and MPC-in-the-Head each defined by distinct mathematical hardness assumptions and representative schemes. This categorization illustrates the diversity of PQC approaches and their varying trade-offs in security, performance, and deployment readiness.}
    \label{fig:pqc-families}
\end{figure}

PQC can be understood as a landscape of families that differ in their underlying hardness assumptions, performance envelopes, and deployment implications. A taxonomy (see~\autoref{fig:pqc-families}) is useful because no single family is uniformly superior across security margins, key and signature sizes, implementation complexity, and resistance to practical attacks. Lattice-based systems currently lead practical standardization due to strong worst-case to average-case reductions and broadly efficient implementations, while code-based and hash-based schemes serve as conservative anchors with long security histories or minimal assumptions. Other families, including multivariate and isogeny-based designs, add diversity and cautionary evidence about structural fragility, whereas MPC-in-the-Head signatures and related symmetric-primitive constructions expand the design space through zero-knowledge techniques and fine-grained engineering trade-offs.

Lattice-based cryptography has emerged as the most versatile family because it offers rigorous reductions and competitive performance across general-purpose central processing units (CPUs), embedded platforms, and accelerators. Security builds on problems such as learning with errors (LWE) and short integer solution, which admit reductions from worst-case lattice problems, giving unusually strong confidence relative to many public-key designs \cite{regev2009lattices}. Module and ring variants preserve these foundations while improving practicality and parameterization, which helps explain the selection of lattice schemes in the first wave of NIST standards for key encapsulation mechanism (KEM) and signatures \cite{langlois2015worst, nist_fips_pqc_2024, nist_csrc_fips_pqc_2024}. Beyond the standards themselves, the ecosystem already includes hardware and algorithmic optimizations for signing and verification, such as low-latency Dilithium pipelines on field-programmable gate arrays (FPGAs), graphics processing unit (GPU) accelerated ML-DSA servers, and Fourier-based engineering for compact signatures in FALCON \cite{10445248, 10748358, fouque2018falcon}. These gains come with real-world challenges. Implementers must handle subtle issues such as discrete Gaussian sampling, rejection behavior, and microarchitectural leakage, and recent work continues to refine both attack surfaces and countermeasures for Kyber and Dilithium style designs \cite{iavich2024investigating, liu2025release, zeitoun2022sidechannel, aikata2022kali}. In short, lattices combine strong theory with a fast-moving engineering front, which is why they are prominent in standards and pilots.

Code-based cryptography occupies a different point in the taxonomy. The security of these schemes rests on the hardness of decoding random linear codes, a problem that has resisted cryptanalysis since the original McEliece proposal in 1978 \cite{mceliece1978public}. This longevity is valuable for risk management because it gives independent assumptions and decades of scrutiny. Modern code-based KEMs such as hamming quasi-cyclic (HQC) continue that lineage while exploring quasi-cyclic structures and implementation techniques that improve practicality \cite{melchor2018hamming, deshpande2023fast}. Work on strengthening and parameterizing McEliece-style systems further illustrates the conservative nature of this family \cite{loidreau2000strengthening}. The principal drawback is well known. Public keys are very large, often by orders of magnitude compared to lattice-based designs, which complicates deployment in bandwidth-sensitive protocols and on constrained devices. For many migration roadmaps, this family remains a security hedge and a domain-specific fit rather than a universal default.

Hash-based signatures provide a second conservative pillar. Their security relies only on the preimage and collision resistance of cryptographic hash functions, which avoids the need to assume the hardness of algebraic structures against quantum adversaries. SPHINCS+ exemplifies this approach by composing one-time signatures, few-time techniques, and hypertrees into a stateless, standardizable framework with careful treatment of the quantum random oracle model \cite{bernstein2019sphincs+}. Engineering work continues to reduce area and energy for embedded deployments and to tune the design of components like forest of random subsets (FORS) for stronger message security guarantees \cite{deshpande2025sphincslet, yehia2020hash}. The trade-offs are pragmatic rather than theoretical. Signatures are large and signing is slower than lattice-based signatures, which affects handshake sizes and throughput in interactive protocols. Still, the minimal assumption set makes hash-based schemes an attractive choice for high assurance applications and long-lived artifacts.

Multivariate cryptography illustrates both innovation and fragility within PQC. These schemes rely on the difficulty of solving systems of multivariate quadratic (MQ) equations over finite fields, which is Nondeterministic Polynomial-time (NP)-hard in general. The family has produced compact signatures and fast signing in some proposals, but it has also faced decisive cryptanalytic pressure. Notable breaks against Rainbow and improved analyses of unbalanced oil and vinegar (UOV) variants have repeatedly reset expectations about safe parameterization \cite{beullens2022breaking, beullens2021improved, kramer2019fault}. Recent work, such as the MAYO multivariate signature scheme (MAYO), revisits design choices and implementation security, including the effect of physical attack models on side-channel robustness \cite{aulbach2024mayo}. The net effect is that multivariate signatures contribute diversity and useful ideas, yet their risk profile remains higher until designs accumulate the same security mileage as lattices or codes \cite{dey2023progress}.

Isogeny-based cryptography offered an appealing promise of sub-kilobyte public keys and signatures, which is compelling for protocols that are highly sensitive to bandwidth and storage. The promise was tempered by cryptanalytic advances that produced efficient key recovery for Supersingular Isogeny Diffie–Hellman (SIDH) and practical breaks for Supersingular Isogeny Key Encapsulation (SIKE). These results drew on deep number theoretic insights and careful exploitation of auxiliary structure, and they materially changed the viability assessment of the family \cite{castryck2023efficient}. Additional work has highlighted side-channel concerns, classical and quantum cryptanalysis in genus 1 and 2, and challenges in key compression and protocol design \cite{de2022zero, weitkamper2023cryptanalysis, naehrig2019dual}. Research continues in this area, but the recent history argues for caution and for treating isogeny schemes as experimental rather than deployment-ready until security evidence matures.

A growing set of signatures constructed from symmetric primitives through zero-knowledge techniques rounds out the taxonomy. These systems, often organized under the MPC in the Head paradigm, derive their security from well-studied symmetric components while replacing number theoretic structure with interactive proof style machinery captured in non-interactive transformations. Recent proposals show meaningful progress. MPC in the Head with Repeated Iterations of Threshold Hashing (MiRitH) achieves multi kilobyte signatures with competitive timings, resource constrained implementations of the Practical Efficient Randomized MPC in the Head with K projections (PERK) reduce memory footprints by orders of magnitude, and new designs from the non structured MQ problem explore the boundary between multivariate assumptions and MPC style proofs \cite{Adj_Barbero_Bellini_Esser_Rivera-Zamarripa_Sanna_Verbel_Zweydinger_2024, Bettaieb_Bidoux_Budroni_Palumbi_Perin_2024, 10629011}. At the same time, the proof systems and encodings introduce distinctive fault and side channel considerations that are now being mapped by the community \cite{mondal2024zkfaultfaultattackanalysis}. These constructions expand the design space and provide assumption diversity, although signature sizes and verification costs still limit where they fit best.

Viewed together, this taxonomy clarifies why standards and pilots have converged on a small set of leading candidates while keeping the broader ecosystem in scope. Lattice-based designs anchor general-purpose deployment through strong reductions and active optimization, code-based and hash-based schemes supply conservative alternatives with long or minimal assumptions, and the remaining families contribute diversity, cautionary evidence, and creative techniques that can influence future rounds. The standards process continues to evolve, including additional signature tracks and status updates that reflect new evidence from cryptanalysis, hardware, and protocol experimentation \cite{alagic2024status, alagic2025status}. A taxonomy that keeps both technical depth and operational realities in view helps practitioners build migration plans that balance performance, interoperability, and risk. The following sections examine each family’s mathematical basis and practical considerations in greater depth.

\subsection{Mathematical Foundations and Security Assumptions}
\label{sec:math-foundation}

PQC schemes depart from the assumptions of classical cryptography by building on mathematical problems that remain resistant to both conventional and quantum algorithms. Among these, lattice-based cryptography has emerged as the most influential paradigm, relying on the hardness of computations within high-dimensional lattices-discrete additive subgroups of Euclidean space that combine rich algebraic structure with strong intractability properties.

\subsubsection{Lattice Problems and Learning With Errors}

A central construct in this area is the LWE problem, introduced by Regev in 2005 \cite{regev2009lattices}. LWE has become the backbone of most efficient lattice-based protocols and is widely regarded as a landmark in the theoretical foundations of PQC. Formally, for a secret vector $\mathbf{s} \in \mathbb{Z}_q^n$ and an error distribution $\chi$ over $\mathbb{Z}_q$, the LWE problem is defined as:

\begin{equation}
\begin{aligned}
\mathrm{LWE}_{n,q,\chi}: \quad &
\text{Distinguish between distributions consisting of} \\
& (\mathbf{a}, \langle \mathbf{a}, \mathbf{s} \rangle + e)
\ \text{and uniformly random pairs} \ 
(\mathbf{a}, u),
\end{aligned}
\end{equation}

where $\mathbf{a} \leftarrow \mathbb{Z}_q^n$, $e \leftarrow \chi$, and $u \leftarrow \mathbb{Z}_q$.

The power of lattice-based systems lies in their worst-case to average-case reductions, which connect cryptographic security to the hardness of solving general lattice problems \cite{micciancio2004almost}. These reductions guarantee that if an adversary can solve random LWE instances, then they could also efficiently solve the most difficult lattice problems in the worst case. This property provides rare, long-term assurances of security, reinforcing confidence in lattice-based primitives that have withstood extensive analysis for over twenty years.

To further improve efficiency while preserving security, the M-LWE problem generalizes LWE by introducing ring structure \cite{langlois2015worst}. For a polynomial ring $R = \mathbb{Z}[X]/(X^n + 1)$ and modulus $q$, M-LWE is defined over the module $R_q^k$ as follows:

\begin{equation}
\begin{aligned}
\mathrm{M\text{-}LWE}_{n,k,q,\chi}: \quad &
\text{Given } (A, \mathbf{b} = A\mathbf{s} + \mathbf{e}) \
\\
& \text{with }  A \in R_q^{k \times k}, \ \mathbf{s} \in R_q^k, \ 
\mathbf{e} \leftarrow \chi^k, \\
& \text{distinguish this distribution from uniform.}
\end{aligned}
\end{equation}

This refinement forms the basis for NIST’s standardized schemes, the module-lattice key encapsulation mechanism (ML-KEM) and the module-lattice digital signature algorithm (ML-DSA). These constructions carefully balance rigorous theoretical guarantees with computational efficiency, making them suitable for practical deployment while inheriting the robust hardness assumptions of classical lattice problems.

\subsubsection{Code-Based Security Foundations}

Code-based cryptography builds upon the computational difficulty of decoding random linear error-correcting codes, a problem that has demonstrated remarkable resilience against algorithmic improvements for over four decades since its introduction by McEliece \citep{mceliece1978public}. The fundamental syndrome decoding problem requires finding low-weight error vectors given syndrome information and random parity-check matrices:

\noindent \textbf{Syndrome Decoding:} Given a matrix $H \in \mathbb{F}_2^{(n-k) \times n}$, a syndrome $s \in \mathbb{F}_2^{n-k}$, and a weight bound $t$, find an error vector $e \in \mathbb{F}_2^n$ such that

\begin{equation}
H e^\top = s^\top \quad \text{and} \quad \mathrm{wt}(e) \leq t.
\end{equation}

This problem remains exponentially hard in the general case despite extensive research in coding theory, information theory, and computational complexity, providing exceptional confidence in long-term security through its deep mathematical foundations and extensive cryptanalytic history. The mathematical foundation of code-based cryptography provides security assurances grounded in decades of theoretical and practical analysis, with the original McEliece cryptosystem continuing to resist attacks despite being subjected to intensive cryptanalytic scrutiny for nearly five decades.

\subsubsection{Hash-Based Cryptographic Foundations}

Hash-based signatures provide the most conservative class of post-quantum primitives, as their security relies solely on the hardness of standard hash function properties. The reduction is both conceptually clear and mathematically rigorous: any successful attack on the signature scheme translates directly into either a preimage or collision attack on the hash function. This reliance on minimal and well-understood assumptions avoids the need for intricate algebraic structures that may later prove susceptible to unforeseen advances in cryptanalysis.

The stateless hash-based incredibly conservative signatures (SPHINCS+) construction embodies decades of refinement in this area. It combines several important innovations, including the use of tweakable hash functions, optimized few-time signature mechanisms, and the FORS technique, which together enable a stateless design with strong provable guarantees \cite{nguyen2019new}. Through its carefully layered design, SPHINCS+ achieves existential unforgeability under chosen-message attacks within the quantum random oracle model, while also addressing the state-management difficulties that limited the practicality of earlier hash-based approaches.

\subsubsection{Multivariate Cryptographic Assumptions}

Multivariate cryptography derives its security from the computational difficulty of solving systems of multivariate polynomial equations over finite fields, a problem whose mathematical structure has been extensively studied in algebraic geometry and computational algebra. The MQ problem provides the foundation for this approach:

\noindent \textbf{MQ Problem:} Given $m$ polynomial equations in $n$ variables over a finite field $\mathbb{F}_q$,
\begin{equation}
p_1(x_1, \ldots, x_n) = p_2(x_1, \ldots, x_n) = \cdots = p_m(x_1, \ldots, x_n) = 0,
\end{equation}
find a solution $\mathbf{x} \in \mathbb{F}_q^n$ that satisfies all equations simultaneously.

While this problem is NP-hard in general, practical multivariate schemes require extraordinarily careful construction to avoid structural vulnerabilities that enable efficient attacks, as demonstrated by the recent cryptanalytic successes against prominent multivariate constructions \cite{beullens2022breaking}.

\section{Core Algorithm Families}
\label{sec:algo-fam}

\subsection{Lattice-Based Cryptography}

\begin{figure}[ht]
    \centering
    \includegraphics[width=0.7\linewidth]{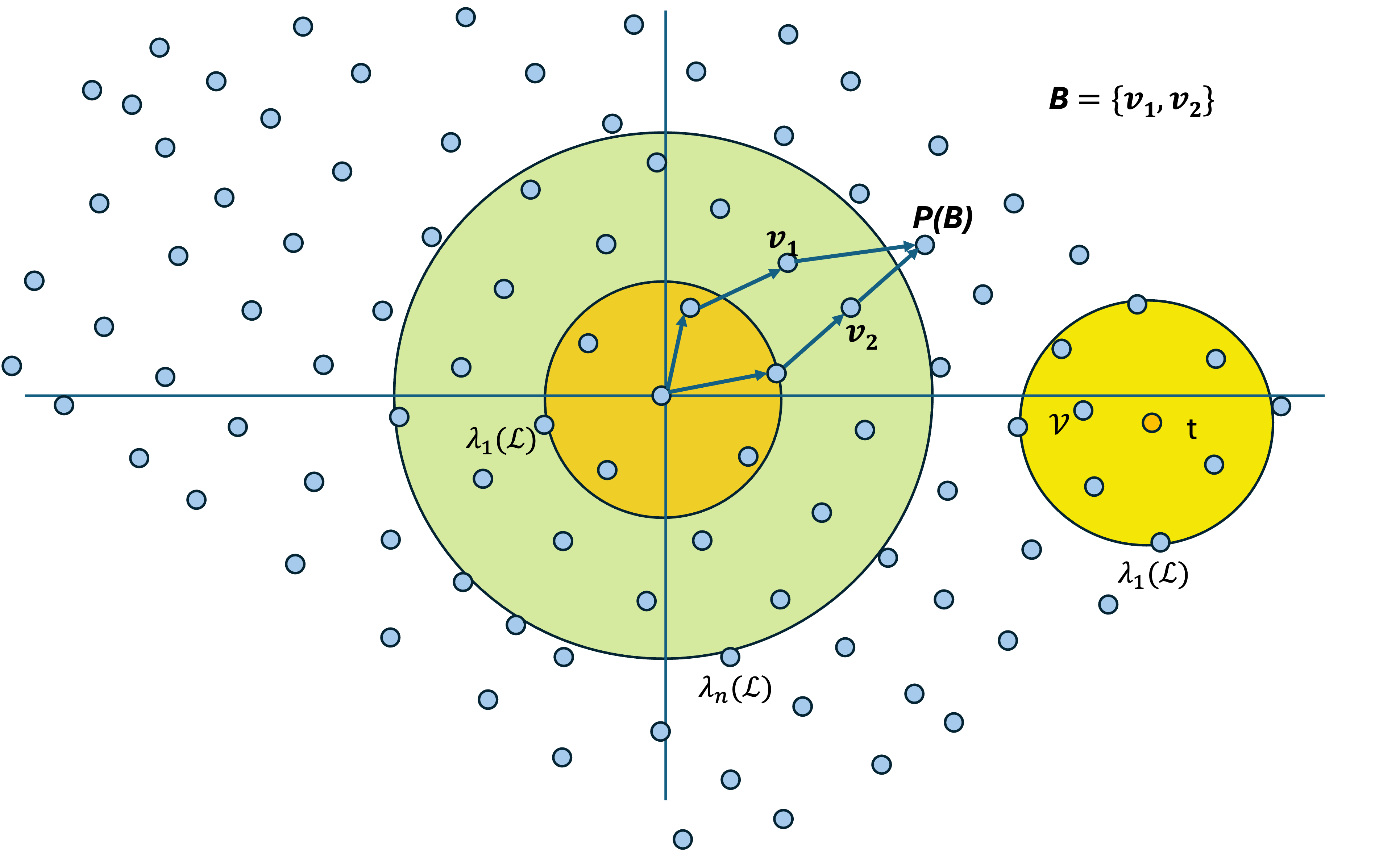}
    \caption{Pictorial representation of a 2D lattice \cite{ravi2021lattice}.}
    \label{fig:2d-lattice}
\end{figure}

Lattice-based cryptographic schemes have established themselves as the preeminent family in post-quantum standardization through their exceptional synthesis of theoretical rigor, practical efficiency, and implementation versatility. The mathematical foundations rest upon the rich structure of lattices in high-dimensional spaces, where the discrete nature of lattice points creates computational problems that remain intractable even for quantum adversaries while enabling efficient cryptographic operations through carefully designed algorithms. Figure~\ref{fig:2d-lattice} illustrates a two-dimensional lattice, where lattice vectors generate a discrete set of points and shortest vector problems emerge naturally from its geometry.

ML-KEM, achieving standardization as FIPS 203, demonstrates the practical viability of lattice-based cryptography through computational efficiency that often surpasses classical alternatives while maintaining strong security guarantees rooted in worst-case lattice problems \cite{cherkaoui2024exploring}. The algorithm operates over polynomial rings with meticulously chosen parameters that balance security requirements against implementation constraints, achieving key encapsulation operations with remarkable efficiency. Performance analyses conducted across diverse hardware platforms indicate that ML-KEM-512 executes approximately three times faster than X25519 elliptic curve Diffie-Hellman while requiring only moderate increases in communication overhead.

ML-DSA, standardized as FIPS 204, employs the mathematically sophisticated Fiat-Shamir with aborts construction that transforms interactive identification protocols into non-interactive signature schemes \cite{devevey2023detailed}. This approach utilizes rejection sampling techniques to ensure that signatures leak no information about the signing key, achieving security reductions to worst-case lattice problems through careful probabilistic analysis. The rejection sampling mechanism represents a crucial innovation that prevents statistical attacks while maintaining computational efficiency, as demonstrated by optimized implementations that complete signing operations in approximately 0.65 milliseconds and verification in 0.53 milliseconds across diverse hardware platforms \cite{demir2025performance}.

FALCON (FN-DSA) offers compact signature alternatives based on Nth-degree Truncated Polynomial Ring Units (NTRU) lattices through the mathematically elegant Gentry-Peikert-Vaikuntanathan framework \cite{fouque2018falcon}. The scheme achieves significantly smaller signature sizes compared to ML-DSA through fast Fourier sampling over NTRU lattices, utilizing advanced mathematical techniques from algebraic number theory to generate signatures with optimal size properties. However, FALCON's implementation complexity, particularly regarding floating-point arithmetic requirements and side-channel resistance, presents additional deployment challenges that must be carefully addressed through specialized implementation techniques.

Recent implementation research has revealed side-channel vulnerabilities in lattice-based implementations, revealing that the mathematical structure that enables efficient computation can also create information leakage channels that compromise security in practical deployments \cite{liu2025release}. Single-trace attacks have demonstrated the possibility of full key recovery using power analysis of number theoretic transform (NTT) operations, while electromagnetic analysis can extract secret keys with minimal traces through sophisticated signal processing techniques. These findings emphasize the critical importance of implementing robust countermeasures, including constant-time operations, masking techniques, and careful compiler optimization management.

\subsection{Code-Based Cryptography}

Code-based cryptographic systems provide the longest-running security foundation among post-quantum families. HQC is a leading code-based KEM that provides algorithmic diversity independent of lattice assumptions, and it remains under active NIST evaluation alongside other code-based proposals \cite{melchor2018hamming,deshpande2023fast,alagic2025status}.

HQC achieves an optimal balance between security and efficiency through its ingenious quasi-cyclic structure, which reduces storage requirements while maintaining the fundamental hardness properties of random linear codes \cite{melchor2018hamming}. The algorithm offers public keys in the practical range of 2-7 kilobytes while maintaining competitive performance characteristics that enable deployment in resource-constrained environments. The selection of HQC over competing code-based schemes reflects NIST's careful evaluation of security analysis maturity, decryption failure rate characterization, and implementation complexity.

Classic McEliece represents the most mathematically conservative code-based approach, having maintained its security properties for over 45 years without experiencing any fundamental cryptanalytic compromises despite intensive research efforts by the international cryptographic community. The scheme offers exceptionally compact ciphertexts and extraordinarily strong security margins through its reliance on well-understood mathematical principles from coding theory, but requires extremely large public keys exceeding one megabyte for high security parameters.

Recent security analysis has refined computational complexity estimates for code-based schemes through sophisticated applications of advanced information set decoding algorithms, which represent the most powerful known attacks against code-based constructions \cite{nguyen2019new}. Low-memory attack scenarios utilizing specialized algorithmic techniques have necessitated parameter adjustments for several constructions, though the fundamental security of well-designed code-based schemes remains intact due to the exponential hardness of the underlying mathematical problems.

\subsection{Hash-Based Signatures}

Hash-based signature schemes provide the most mathematically conservative security foundation in PQC, relying exclusively on hash function security rather than complex mathematical assumptions that might be vulnerable to future cryptanalytic advances or unexpected algorithmic breakthroughs. SPHINCS+ (SLH-DSA), achieving standardization as FIPS 205, represents the culmination of decades of theoretical and practical research in stateless hash-based constructions \cite{nguyen2019new}.

The SPHINCS+ framework incorporates sophisticated mathematical innovations, including tweakable hash functions that provide domain separation and prevent certain classes of attacks, optimized few-time signature schemes that minimize computational overhead, and FORS constructions that enable efficient authentication of large message spaces while maintaining stateless operation. Unlike earlier hash-based schemes that required careful state management to prevent catastrophic key reuse vulnerabilities, SPHINCS+ enables unlimited signature generation without security degradation through its mathematically elegant stateless design.

Performance optimization efforts have achieved substantial improvements in hash-based signature efficiency through specialized implementations that exploit the inherent parallelism in hash-based constructions and algorithmic enhancements that reduce computational overhead. Hardware accelerations utilizing advanced instruction sets demonstrate meaningful performance improvements over reference implementations, though hash-based signatures continue producing substantially larger signatures than lattice-based alternatives, typically ranging from 8-30 kilobytes depending on security parameters and optimization strategies.

\subsection{Multivariate Cryptography}

Multivariate cryptographic schemes have experienced profound challenges due to sophisticated algebraic attacks that exploit the rich mathematical structure underlying these constructions, fundamentally altering the landscape for this algorithmic family and raising questions about the long-term viability of multivariate approaches in PQC. The cryptanalytic breakthrough against Rainbow, demonstrated through the development of rectangular MinRank attacks, represents a watershed moment in multivariate cryptography \cite{beullens2022breaking}.

The UOV construction serves as the mathematical foundation for most contemporary multivariate schemes, utilizing a sophisticated trapdoor structure that enables efficient signing operations while maintaining the apparent intractability of the public polynomial system for potential adversaries. However, the rectangular MinRank attack developed against Rainbow has demonstrated applicability to other multivariate constructions, revealing fundamental vulnerabilities in the mathematical approaches that underlie this family of schemes \cite{kramer2019fault}.

Recent cryptanalytic analysis indicates that variants of sophisticated algebraic attacks may compromise parameter sets with computational complexity as low as $2^{55}$ operations, representing a dramatic reduction from the security levels originally claimed for these constructions. These attacks exploit subtle mathematical relationships in the polynomial structure that were not apparent during initial security analysis, demonstrating how the complex algebraic structure that enables efficient multivariate operations can also create vulnerabilities that become apparent only through sophisticated cryptanalytic techniques.

\subsection{Isogeny-Based Cryptography}

The isogeny-based approach to PQC suffered a catastrophic and unexpected setback with the July 2022 cryptanalytic breakthrough against SIDH/SIKE, which demonstrated how seemingly secure mathematical constructions can harbor subtle vulnerabilities that become apparent only through sophisticated attacks utilizing advanced mathematical techniques \cite{de2022zero}. The attack exploits auxiliary point information included in the protocol specification, enabling classical key recovery in approximately one hour on standard hardware through mathematical techniques that do not require quantum computation.

SIKE initially appeared to be a highly promising PQC candidate because of its exceptionally small key sizes, with public keys remaining under 600 bytes even at strong security levels \cite{naehrig2019dual}. Its efficiency in bandwidth consumption, coupled with security assumptions rooted in the hardness of computing isogenies between supersingular elliptic curves, positioned it as one of the leading contenders during the NIST post-quantum standardization process.

The eventual cryptanalytic attack on SIKE showcased striking mathematical depth, drawing on advanced genus theory and the little-known "glue-and-split" theorem \cite{weitkamper2023cryptanalysis}. This result underscored how powerful breakthroughs can arise from unexpected areas of mathematics, revealing vulnerabilities that had not been captured in the original security analysis of isogeny-based cryptography. The episode highlights a broader lesson for PQC: security guarantees rest on unproven hardness assumptions that may conceal subtle weaknesses until uncovered by future advances.

\subsection{MPC-in-the-Head Approaches}

The MPC-in-the-Head paradigm represents an innovative and mathematically sophisticated approach to constructing post-quantum signature schemes through zero-knowledge proof techniques that transform arbitrary symmetric cryptographic primitives into quantum-resistant signature schemes while providing exceptional flexibility in underlying security assumptions \cite{nguyen2019new}. This framework enables cryptographic constructions based on well-studied symmetric primitives such as block ciphers and hash functions, potentially offering security guarantees that inherit the strength of these thoroughly analyzed building blocks.

Syndrome Decoding in the Head represents a particularly promising variant of the MPC-in-the-Head approach, achieving more compact signatures approaching 8-17 kilobytes through sophisticated mathematical constructions that optimize the zero-knowledge proof system for the specific structure of syndrome decoding problems \cite{nguyen2019new}. This approach demonstrates how careful mathematical analysis can identify opportunities for optimization within the general MPC-in-the-Head framework.

Despite theoretical advantages and recent efficiency improvements, MPC-in-the-Head constructions continue to produce substantially larger signatures than alternative post-quantum approaches, often ranging from tens to hundreds of kilobytes, depending on the specific construction and security parameters. This overhead stems from the fundamental mathematical structure of zero-knowledge proofs, which must include multiple proof components, commitment values, and verification information to enable recipients to verify the proof while preventing forgery attacks.

\section{Digital Signature Schemes}
\label{sec:digital-schemes}

Quantum-resistant digital signatures are a cornerstone for ensuring authenticity in diverse applications such as software distribution, transport layer security (TLS) certificates, and blockchain systems. To this end, NIST has endorsed three signature families for widespread adoption-ML-DSA (Dilithium), SLH-DSA (SPHINCS+), and the lattice-based Falcon under a parallel track-highlighting a balance between performance efficiency and conservative security across lattice- and hash-based paradigms \cite{nist_fips_pqc_2024,nist_csrc_fips_pqc_2024,alagic2024status}.

These post-quantum signature mechanisms mark a decisive move away from traditional RSA and elliptic curve digital signature algorithm (ECDSA), grounding security in problems that remain intractable even with quantum computation \cite{liu2024post}. The selected algorithms-CRYSTALS-Dilithium (ML-DSA), Falcon (FN-DSA), and SPHINCS+ (SLH-DSA)-each bring unique trade-offs in efficiency, key and signature sizes, and assurance levels, ensuring flexibility for varied deployment requirements in the emerging quantum era \cite{commey2025performance}.

\subsection{CRYSTALS-Dilithium (ML-DSA)}
\label{subsubsec:crystals-dilithium-ml-dsa}

\begin{figure}
    \centering
    \includegraphics[width=0.96\linewidth]{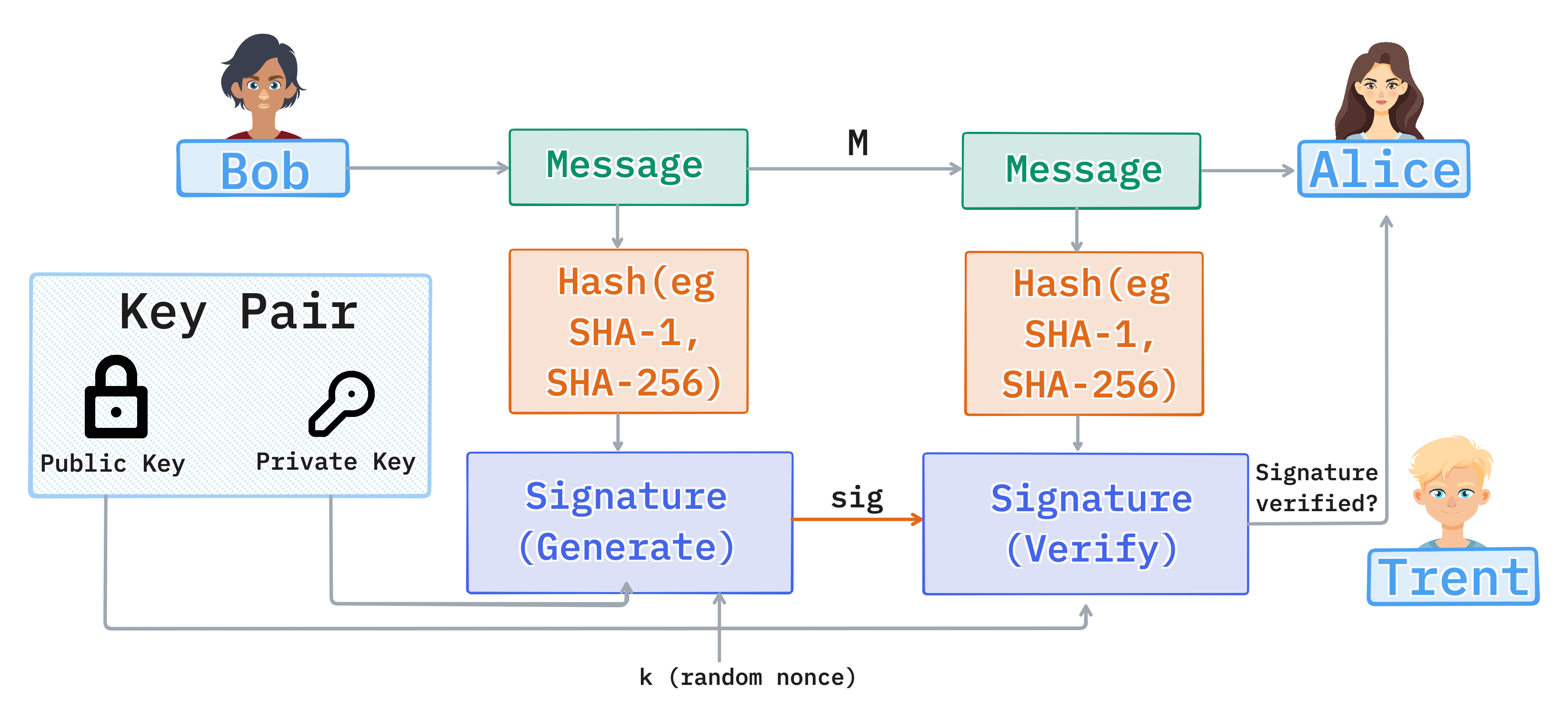}
    \caption{The ML-DSA (CRYSTALS-Dilithium) signature process. 
    Bob signs a message $M$ by applying his private key to the message hash, generating a signature $(\text{sig})$. 
    He transmits $(M, \text{sig})$ to Alice, who verifies integrity and authenticity using Bob’s public key and the hash of $M$. 
    A trusted authority, Trent, may participate in key pair certification or management \cite{asecuritysite_21337}.}
    \label{fig:ml-dsa}
\end{figure}

The ML-DSA (see~\autoref{fig:ml-dsa}), formerly known as CRYSTALS-Dilithium, is a lattice-based signature scheme standardized in FIPS 204. Its security relies on the hardness of the module short integer solution (MSIS) and module learning with errors (MLWE) problems \cite{goertzen2022postquantumsignaturesdnssecrequestbased}. ML-DSA is particularly effective in constrained environments, achieving faster signing speeds than its competitors-0.65 ms compared to Falcon (3.28 ms) and SPHINCS+ (131.9 ms)-at NIST Security Level 2. Although its public keys (2,592 bytes) and signatures (3,309 bytes) are larger than those of pre-quantum schemes, ML-DSA strikes a balance between efficiency and scalability, making it well-suited for real-time applications such as blockchain-based federated learning \cite{Li_2024, commey2025pqs}. 

Active research continues on PQC integration with domain name system security extensions (DNSSEC), including fragmentation strategies for large signatures, though PQ-based signatures are not yet widely deployed in production DNSSEC \cite{goertzen2022postquantumsignaturesdnssecrequestbased}. 

\subsection{Falcon (FN-DSA)}
\label{subsubsec:falcon-fn-dsa}

The FN-DSA (Falcon) scheme is an NTRU-lattice-based construction that leverages Fast Fourier sampling to achieve compact signatures. Tracked in NIST’s additional digital signature process, Falcon complements ML-DSA and SLH-DSA as a candidate with distinct efficiency and size trade-offs \cite{alagic2024status,fouque2018falcon}. Its design follows a ``hash-and-sign'' paradigm, where a short vector $\mathbf{v}$ is computed in the lattice defined by the public key $\mathbf{A}$ and the hash of a message $H(m)$, such that $\mathbf{A} \cdot \mathbf{v} = H(m)$ \cite{Li_2024}. This construction yields the smallest signature sizes among NIST finalists, with Falcon-512 signatures measuring only 666 bytes, making it attractive for bandwidth-sensitive applications such as DNSSEC \cite{goertzen2022postquantumsignaturesdnssecrequestbased}.

Performance benchmarks show Falcon’s strength in verification, with speeds near 0.3 ms due to fast fourier transform (FFT) based optimizations \cite{Li_2024}. However, key generation ($\sim$5.4 ms) and signing ($\sim$3.28 ms) remain computationally demanding because of complex Gaussian sampling, creating challenges for lightweight devices \cite{commey2025pqs}. Hardware accelerators such as FPGA or application-specific integrated circuit (ASIC) implementations can reduce this overhead, though they may introduce side-channel vulnerabilities \cite{goertzen2022postquantumsignaturesdnssecrequestbased}. 

While Falcon’s compact signatures make it well-suited for DNSSEC, where user datagram protocol (UDP) packet size constraints strongly favor smaller payloads, its slower signing speed limits usability in latency-critical scenarios such as federated learning \cite{commey2025pqs}. Moreover, reliance on the NTRU assumption-less extensively analyzed than MLWE-leaves open long-term theoretical questions, and experts caution that rejection-sampling techniques may weaken under real-world side-channel conditions \cite{Li_2024}.

\subsection{SPHINCS+ (SLH-DSA)}
\label{subsubsec:sphincs-slh-dsa}

\begin{figure}
    \centering
    \includegraphics[width=0.96\linewidth]{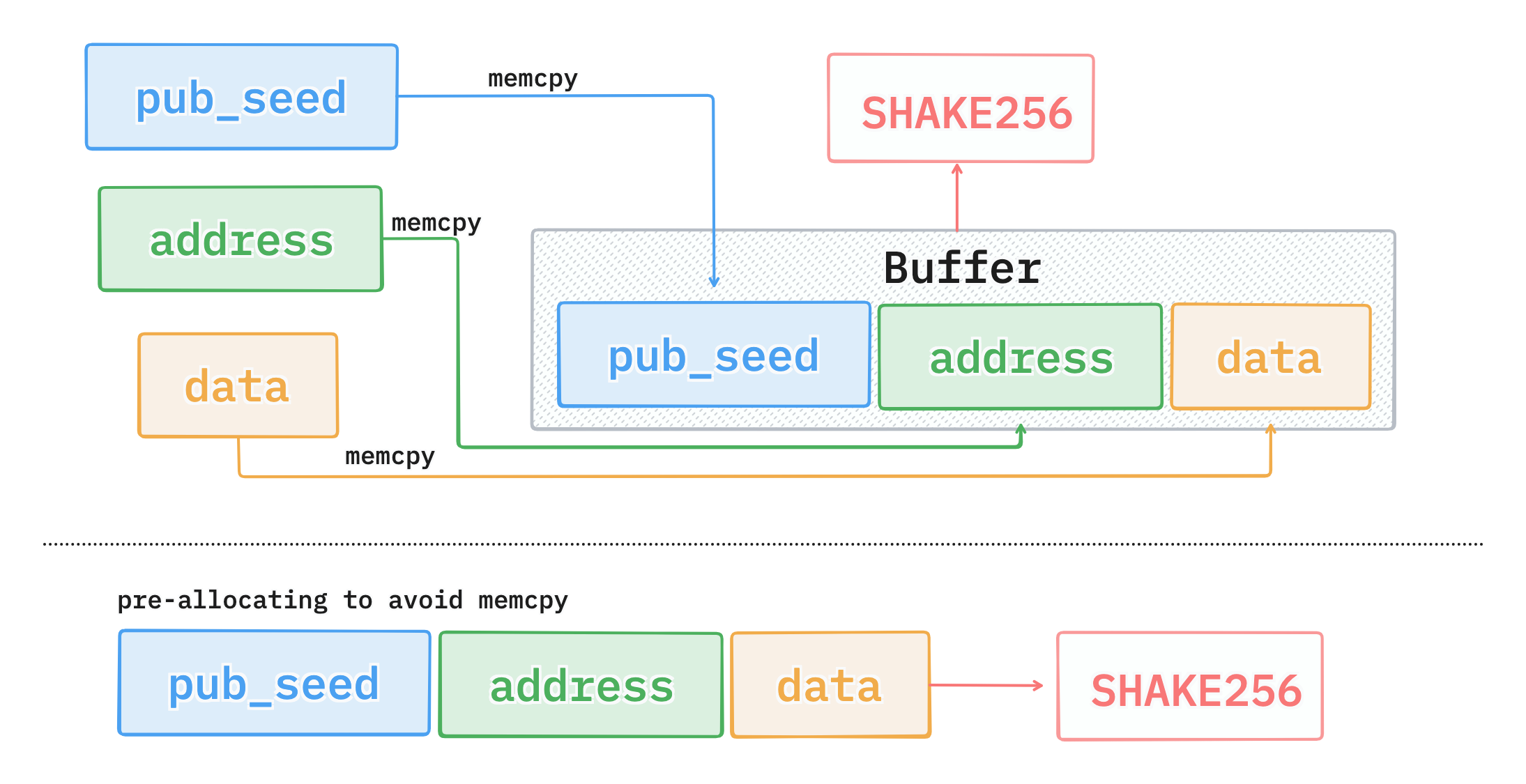}
    \caption{Illustration of SLH-DSA (SPHINCS+) input processing for SHAKE256. 
The top approach shows a buffer-based method where \texttt{pub\_seed}, \texttt{address}, and \texttt{data} are copied into a buffer before hashing, incurring multiple \texttt{memcpy} operations. 
The bottom approach demonstrates a pre-allocation strategy that directly arranges inputs for SHAKE256, avoiding redundant memory copies and improving efficiency \cite{ye2025rvslh}.}
    \label{fig:slh-dsa}
\end{figure}

The stateless hash-based digital signature algorithm (SLH-DSA), instantiated through SPHINCS+, has recently been standardized as FIPS 205 (see~\autoref{fig:slh-dsa}) \cite{alagic2024status}. Unlike lattice-based approaches, SLH-DSA derives its security solely from the collision resistance of hash functions, avoiding reliance on newer hardness assumptions. It achieves quantum resistance through a hierarchical structure of hash trees combined with few-time signature mechanisms. Specifically, winternitz one-time signature plus (WOTS+) is used for signing individual nodes, FORS enables efficient few-time signatures at the leaf level, and the Hypertree construction links these components together to achieve stateless operation \cite{deshpande2025sphincslet}.

The scheme’s performance depends heavily on the chosen parameter set. For instance, SLH-DSA-128s (128-bit security, small variant) produces signatures of 7,856 bytes, with signing times near 131~ms and verification around 3.6~ms \cite{commey2025pqs, Li_2024}. Both Secure Hash Algorithm (SHA-2, specifically SHA-256 and SHA-512) and Secure Hash Algorithm Keccak (SHAKE256) are supported as underlying primitives, with SHA-2 generally providing up to $2\times$ faster signing performance on hardware due to existing optimizations \cite{deshpande2025sphincslet}.

Recent hardware implementations such as SPHINCSLET further enhance efficiency. For SHAKE256-based SLH-DSA-128s, FPGA designs on Artix-7 achieve area usage of only 10.8K look-up tables (LUTs), while delivering $2.5$--$5\times$ higher throughput compared to software-assisted approaches \cite{deshpande2025sphincslet}. SHA-2 variants demonstrate even greater improvements, with reported $2$--$4\times$ speedups across security levels. While SLH-DSA has relatively large signatures and slower signing compared to lattice-based alternatives, its reliance on conservative and well-understood hash function properties makes it an appealing option for long-term quantum resilience \cite{alagic2024status}.

\subsection{NIST Additional Digital Signature Candidates}
\label{subsec:additional-sigs}

To improve design diversity and hedge against correlated assumptions, NIST reopened its call for additional post-quantum digital signature schemes. NIST Interagency Report (NIST IR)~8528 summarizes the status and goals of this track and motivates a broader portfolio beyond the first three FIPS approvals \cite{alagic2024status,alagic2025status}. Table~\ref{tab:additional-sigs} consolidates the candidate set into a single view. We list each scheme, the underlying family, a short design note, and the primary references available in our bibliography. When a scheme-specific paper is outside our current bibliography, we cite the NIST IR for authoritative program context.

\begin{table}[t]
\centering
\small
\caption{NIST additional digital signature candidates: families, design notes, and references. Where a scheme reference is not present in our bibliography, we cite the NIST IR \cite{alagic2024status}.}
\label{tab:additional-sigs}
\begin{tabular}{p{3cm} p{2.5cm} p{7.3cm}}
\toprule
    \rowcolor{tabheader}%
    \textcolor{white}{\textbf{Scheme (Refs)}} &
    \textcolor{white}{\textbf{Family}} &
    \textcolor{white}{\textbf{Design note}}
    \\
\midrule
CROSS~\cite{alagic2024status} & Multivariate & Non-structured MQ design, targets compact signatures with careful soundness analysis \\
LESS~\cite{alagic2024status} & Multivariate & MQ-based, explores parameter sets for practical verification costs \\
HAWK~\cite{alagic2024status} & Multivariate & MQ approach with attention to implementation footprint for constrained devices \\
MiRitH~\cite{Adj_Barbero_Bellini_Esser_Rivera-Zamarripa_Sanna_Verbel_Zweydinger_2024} & MPC-in-the-Head & MinRank-in-the-Head, improved encodings and security analysis, multi-kilobyte signatures with competitive timings \\
PERK~\cite{Bettaieb_Bidoux_Budroni_Palumbi_Perin_2024} & MPC-in-the-Head & MPCitH family, memory-reduction techniques enable Arm Cortex M-class implementations and large footprint cuts \\
RYDE~\cite{alagic2024status} & MPC-in-the-Head & MPCitH design focusing on round and memory complexity \\
SDitH~\cite{mondal2024zkfaultfaultattackanalysis,alagic2024status} & MPC-in-the-Head & Symmetric-primitive based MPCitH signature; fault and side-channel considerations under active attacks \\
MQOM~\cite{10629011} & Multivariate & “MQ on my Mind,” non-structured MQ with MPC-style techniques for practical signatures \\
UOV~\cite{beullens2021improved,alagic2024status} & Multivariate & UOV-style constructions and modern cryptanalysis inform conservative parameter choices \\
MAYO~\cite{aulbach2024mayo,alagic2024status} & Multivariate & Recent optimizations and implementation-security evaluation under physical attack models \\
QR-UOV~\cite{alagic2024status} & Multivariate & UOV variant with quasi-random structure, explores trade-offs in key size and verification \\
SNOVA~\cite{alagic2024status} & Multivariate & Structured MQ approach aiming for smaller signatures and faster verification \\
FAEST~\cite{alagic2024status} & Symmetric-based & VOLE-in-the-Head signatures from symmetric primitives, avoids algebraic structure \\
SQIsign~\cite{alagic2024status,castryck2023efficient,weitkamper2023cryptanalysis} & Isogeny & Isogeny-based signatures pursuing very compact artifacts, considered with caution given recent breaks in related primitives \\
\bottomrule
\end{tabular}
\end{table}

\paragraph{Context and takeaways.}
The additional-signatures effort explicitly seeks assumption diversity and complementary performance profiles \cite{alagic2024status}. MPC-in-the-Head designs (MiRitH, PERK, SDitH) leverage symmetric primitives and zero-knowledge encodings, which improves assumption diversity but increases signature size and memory pressure \cite{Adj_Barbero_Bellini_Esser_Rivera-Zamarripa_Sanna_Verbel_Zweydinger_2024,Bettaieb_Bidoux_Budroni_Palumbi_Perin_2024}. Non-structured multivariate schemes (Multivariate Quadratic Oil and Mayonnaise (MQOM), UOV variants, MAYO) continue to evolve under active cryptanalysis \cite{beullens2021improved,aulbach2024mayo}. Isogeny-based approaches such as Supersingular Quaternion Isogeny Signatures (SQIsign) target extreme compactness, and they must be weighed against recent lessons from SIDH and SIKE attacks \cite{castryck2023efficient,weitkamper2023cryptanalysis}. This portfolio underscores that future standardization should consider not only speed and sizes, but also assumption independence and implementation security.

\section{Performance Evaluation}
\label{sec:perf-eval}

The comprehensive performance analysis (see~\autoref{tab:perf-ana}) of PQC algorithms reveals distinct computational and communication trade-offs among algorithm families that fundamentally determine their suitability for different deployment scenarios and application requirements. Lattice-based algorithms achieve an optimal balance across multiple performance criteria, offering excellent computational efficiency, manageable communication overhead, and strong theoretical security foundations that combine to make them suitable for widespread deployment across diverse computing environments.

\begin{table}[ht]
\caption{Post-Quantum Algorithm Performance Analysis}
\small
\centering
\label{tab:perf-ana}
\begin{tabular}{lccccccc}
\toprule
\rowcolor{tabheader}%
    \textcolor{white}{\textbf{Algorithm}} &
    \textcolor{white}{\textbf{Sec. Level}} &
    \textcolor{white}{\textbf{KeyGen}} &
    \textcolor{white}{\textbf{Enc/Sign}} &
    \textcolor{white}{\textbf{Dec/Ver}} &
    \textcolor{white}{\textbf{PK (B)}} &
    \textcolor{white}{\textbf{Sig/CT (B)}}
    \\
\midrule
\textbf{ML-KEM-512} \cite{demir2025performance} & L1 & 0.253 & 0.070 & 0.084 & 800 & 768 \\
\textbf{ML-KEM-768} \cite{demir2025performance} & L3 & 0.354 & 0.095 & 0.118 & 1,184 & 1,088 \\
\textbf{ML-KEM-1024} \cite{demir2025performance} & L5 & 0.483 & 0.138 & 0.159 & 1,568 & 1,568 \\
\textbf{ML-DSA-44} \cite{demir2025performance} & L1 & 0.253 & 0.840 & 0.267 & 1,312 & 2,420 \\
\textbf{ML-DSA-65} \cite{demir2025performance} & L3 & 0.392 & 1.205 & 0.398 & 1,952 & 3,293 \\
\textbf{ML-DSA-87} \cite{demir2025performance} & L5 & 0.652 & 1.998 & 0.584 & 2,592 & 4,595 \\
\textbf{FALCON-512} \cite{fouque2018falcon} & L1 & 8.64 & 0.168 & 0.036 & 897 & 666 \\
\textbf{FALCON-1024} \cite{fouque2018falcon} & L5 & 27.45 & 0.344 & 0.073 & 1,793 & 1,280 \\
\textbf{SLH-DSA-128s} \cite{nguyen2019new} & L1 & 0.032 & 120.5 & 1.45 & 32 & 7,856 \\
\textbf{SLH-DSA-192s} \cite{nguyen2019new} & L3 & 0.048 & 285.2 & 2.98 & 48 & 16,224 \\
\textbf{SLH-DSA-256s} \cite{nguyen2019new} & L5 & 0.064 & 652.8 & 5.12 & 64 & 29,792 \\
\textbf{HQC-128} \cite{deshpande2023fast} & L1 & 2.84 & 4.12 & 8.95 & 2,249 & 4,433 \\
\textbf{HQC-192} \cite{deshpande2023fast} & L3 & 5.67 & 8.34 & 18.2 & 4,522 & 8,978 \\
\textbf{HQC-256} \cite{deshpande2023fast} & L5 & 11.2 & 16.8 & 35.4 & 7,245 & 14,421 \\
\bottomrule
\end{tabular}
\begin{flushleft}
\footnotesize \textit{Note: KeyGen, Enc/Sign, Dec/Ver in ms..}
\end{flushleft}
\end{table}

The performance analysis demonstrates that ML-KEM achieves remarkable computational efficiency with sub-millisecond operation times across all security levels, reflecting the mathematical elegance of lattice-based constructions and their compatibility with modern hardware architectures. ML-DSA provides competitive signing performance with verification times remaining under 1.2 milliseconds even for the highest security parameters, demonstrating that post-quantum digital signatures can achieve computational efficiency comparable to or exceeding classical alternatives.

Hash-based signatures exhibit fundamentally different performance characteristics that reflect their conservative security approach and mathematical structure. SLH-DSA signing operations require substantially longer execution times, ranging from 120ms to 650ms, depending on the security level, primarily due to the complex tree-based computations and multiple hash evaluations required for each signature generation.

\subsection{Hardware Acceleration and Implementation Optimization}

Hardware acceleration represents a critical enabler for practical PQC deployment, with lattice-based schemes demonstrating exceptional potential for performance improvements through specialized hardware implementations and optimized software techniques \cite{deshpande2023fast}. Advanced Vector Extensions 2 (AVX2) vector instruction optimizations achieve performance improvements of 3-6$\times$ across all lattice-based algorithms, exploiting the parallelism inherent in polynomial arithmetic operations and matrix computations that form the mathematical foundation of these schemes.

FPGA implementations have achieved remarkable efficiency improvements through specialized architectures that optimize the mathematical operations fundamental to post-quantum algorithms \cite{deshpande2023fast}. Custom polynomial arithmetic units designed specifically for lattice-based operations can achieve substantial performance gains compared to general-purpose processors, while maintaining the flexibility necessary to support multiple algorithms and parameter sets.

\subsection{Network Performance Impact and Deployment Considerations}

Evaluating network performance is essential for understanding how post-quantum algorithms affect communication protocols, bandwidth demands, and deployment feasibility in real-world systems \cite{paquin2020benchmarking}. Lattice-based schemes generally impose limited overhead, with TLS handshake costs increasing by less than 35\% compared to classical counterparts. Their bandwidth impact is moderate and considered acceptable for most application domains. 

However, studies show that when packet loss exceeds 3-5\%, algorithms that rely on fragmenting large messages across multiple packets experience noticeable degradation in performance \cite{paquin2020benchmarking}. Hash-based signature schemes are particularly challenging in this regard: their large signature sizes lead to handshake overheads ranging from 245\% to 890\% relative to classical algorithms, primarily due to the need to transmit signatures over multiple packets.

\subsection{Implementation Security and Side-Channel Considerations}

While post-quantum algorithms offer strong theoretical guarantees, their practical security critically depends on robust implementations. Side-channel attacks-exploiting power consumption, electromagnetic emissions, timing variations, and similar physical leakages-pose a significant risk to PQC deployments \cite{zeitoun2022sidechannel}. Empirical assessments such as test vector leakage assessment (TVLA) have revealed exploitable leakage across several PQC families, underscoring the need for comprehensive mitigation strategies. 

Developing countermeasures requires a trade-off between security strength and computational cost. Lightweight protections are often insufficient, as advanced adversaries have demonstrated key recovery using thousands to hundreds of thousands of observations, depending on the target algorithm and attack vector \cite{liu2025release}. Effective defenses, therefore, demand carefully engineered, multi-layered countermeasures, even though these may introduce notable performance penalties.

\section{System-Level Considerations}
\label{sec:sys-lvl-considerations}

The transition to PQC is not only an algorithmic choice but a full-stack engineering program that touches protocols, infrastructure, operations, and governance. The first three NIST FIPS approvals for ML-KEM, ML-DSA, and SLH-DSA formalized a baseline for federal and commercial deployments, and ongoing NIST status reports and additional tracks continue to adjust priorities as new evidence accumulates \cite{nist_fips_pqc_2024, nist_csrc_fips_pqc_2024, alagic2025status, alagic2024status}. In parallel, the Commercial National Security Algorithm Suite 2.0 and industry guidance signal a clear policy direction that favors crypto-agility and phased rollout rather than single-shot replacement \cite{utimaco2024CNSA2}. The practical meaning is that system owners must plan for mixed environments where classical and post-quantum mechanisms coexist, where version negotiation has to be resilient to downgrade, and where monitoring tracks both performance and security regressions during migration. NIST’s National Cybersecurity Center of Excellence (NCCoE) emphasizes crypto-agility as a programmatic capability, which pushes teams to build inventories, abstract cryptographic dependencies, and design rapid rotation procedures for keys and algorithms \cite{nist2023cryptoAgility}.

\subsection{Protocol Integration}
Protocol integration remains the most visible source of friction in PQC migration. Post-quantum KEMs and signatures change wire images, certificate chains, handshake sizes, and retry behavior. Public deployments and measurements show that the overhead is manageable for lattice KEMs in typical web settings, though it varies across stacks and networks \cite{cloudflare2024state, paquin2020benchmarking}. Signature-heavy handshakes stress bandwidth and path maximum transmission unit (MTU), potentially triggering fragmentation or failure in middleboxes that do not expect larger records. Operators have explored hybrid approaches such as key encapsulation mechanism-based TLS (KEMTLS), which replaces handshake signatures with KEM-based authentication to reduce certificate bloat and mitigate path issues \cite{schwabe2021kemtls}. Similarly, TLS 1.3 hybrids combining classical key exchange with ML-KEM improve near-term resiliency but increase endpoint and public key infrastructure (PKI) complexity \cite{garcia2023quantum, giron2023hybrid}. 

The PKI layer itself demands special attention. Certificate chains with post-quantum signatures are larger, online certificate status protocol (OCSP) responses grow in size, and caches warm more slowly. Operational studies document real-world drawbacks in TLS environments and recommend profiling handshake fragmentation, content delivery network (CDN) edge behavior, and certificate rollover procedures before broad enablement \cite{zacharopoulos2024drawbacks, cloudflare2024state}. Similar considerations apply to DNSSEC, where post-quantum signatures may exceed typical UDP response limits, motivating request-based fragmentation and adaptive transport selection to prevent resolution failures \cite{goertzen2022postquantumsignaturesdnssecrequestbased}.

\subsection{Embedded and Implementation Constraints}
Constrained and heterogeneous devices add another dimension to PQC adoption. IoT gateways, sensors, and embedded controllers have strict limits on code size, random-access memory (RAM), energy, and update bandwidth, which makes key sizes, stack depth, and constant-time countermeasures primary design constraints rather than post-hoc optimizations~\cite{liu2024post, gsma2025PQIoT, fitzgibbon2024constrained}. Insights from recent TinyML and edge-AI studies highlight that similar constraints arise in low-power inference workloads — hardware-aware co-design, quantization, and instruction scheduling directly affect cryptographic feasibility~\cite{somvanshi2025tiny}.
 Experimental integrations of PQC into constrained application protocol (CoAP) and message queuing telemetry transport for sensor networks (MQTT-SN) confirm that message sizes and handshake round-trip times must be tuned through transport parameters, session resumption, and caching to sustain reliability at scale~\cite{blanco2024integrating}. For long-lived deployments, crypto-agility becomes a lifecycle obligation, requiring firmware and bootloaders to accept new trust anchors and algorithms, stable abstraction layers for hardware security module (HSM) interfaces, and remote attestation formats that accommodate larger evidence objects. Sector-specific guidelines in banking and critical infrastructure emphasize that migration must extend beyond libraries to include network resiliency, operational playbooks, and auditability~\cite{bettale2022postquantum, pape2024beyondPQC, auer2023projectLeap, del2024cybersecurity}.

Implementation security further reinforces this system-level view. Most real-world vulnerabilities stem from leakage or fault behavior rather than cryptanalysis. Side-channel hardening for lattice schemes requires attention to discrete Gaussian sampling, masking, and constant-time polynomial arithmetic. Prior work has categorized leakage classes, built countermeasures, and shown that naive optimizations can reintroduce risk~\cite{zeitoun2022sidechannel, iavich2024investigating, liu2025release}. Hardware acceleration bridges performance gaps while maintaining constant-time guarantees, with FPGA and GPU designs for ML-DSA, FFT and discrete Gaussian co-designs for FALCON, and compact accelerators for SPHINCS+~\cite{10445248, 10748358, fouque2018falcon, karabulut2024hardware, alsuhli2024area, deshpande2025sphincslet, lee2024high}. Secure deployment requires coupling such accelerators with compiler fences, microarchitectural isolation, and continuous test harnesses for fault injection, power, and EM analysis, and randomized protocol paths.

\subsection{Operational Readiness and Ecosystem Maturity}
Operational measurement and readiness are essential for sustaining confidence during PQC deployment. Internet-scale telemetry and controlled benchmarks have become critical for tracking adoption rates, failure signatures, and the effectiveness of hybrid designs in production environments \cite{sowa2024post}. Cloud and web-scale studies of TLS handshakes with PQC provide data for rollout schedules, buffer sizing, and retry logic, while identifying where middleboxes or legacy clients require remediation \cite{cloudflare2024state, paquin2020benchmarking}. In finance and payments, pilots and guidance documents from central bank digital currency research and enterprise security programs recommend staged enablement, with emphasis on HSMs, custody workflows, and stress-tested cutover simulations \cite{nili2024cbdcQuantum, auer2023projectLeap, pape2024beyondPQC}. Similar transitions are observed in automotive and transportation networks where vehicle-to-everything (V2X) systems must handle larger credentials and verification delays without compromising safety properties \cite{verma2024quantum, hsu2024vehicle}. 

Interoperability with quantum communications introduces new architectural opportunities. Experiments have demonstrated that quantum key distribution (QKD) can be integrated with PQC in hybrid stacks, combining optical key freshness with classical authenticity \cite{wang2021experimental, garms2024experimental}. Such hybrid architectures are promising for specialized networks where physical layer control is feasible, while broader internet deployments continue to rely on software-based PQC with crypto-agility \cite{garg2024post}. 

A credible system-level plan inventories all cryptographic uses, prioritizes high-value assets, stages hybrid rollouts with rollback options, and aligns with FIPS and CNSA 2.0 milestones. It applies NCCoE guidance for agility, uses telemetry-driven feedback loops, and treats PQC as a continuous capability rather than a one-time upgrade \cite{nist_fips_pqc_2024, nist_csrc_fips_pqc_2024, utimaco2024CNSA2, nist2023cryptoAgility, sowa2024post, cloudflare2024state}. The most mature programs budget for periodic parameter updates, coordinate PKI and protocol teams, and invest in reproducible benchmarking that covers both cryptographic kernels and end-to-end user experience. This holistic approach turns strong algorithms into dependable, measurable, and sustainable services.

\section{Domain-Specific Implications}
\label{sec:domain-spec}

The transition to PQC will not unfold uniformly across industries. Different domains face distinct operational constraints, regulatory drivers, and risk tolerances that shape both the urgency and feasibility of migration. This section analyzes the implications of PQC adoption across the web ecosystem, IoT and embedded devices, financial services, blockchain systems, cloud environments, and the broader regulatory landscape. We conclude with lessons learned that cut across domains.

\subsection{Web and Internet Protocols}
\begin{figure}
    \centering
    \includegraphics[width=\linewidth]{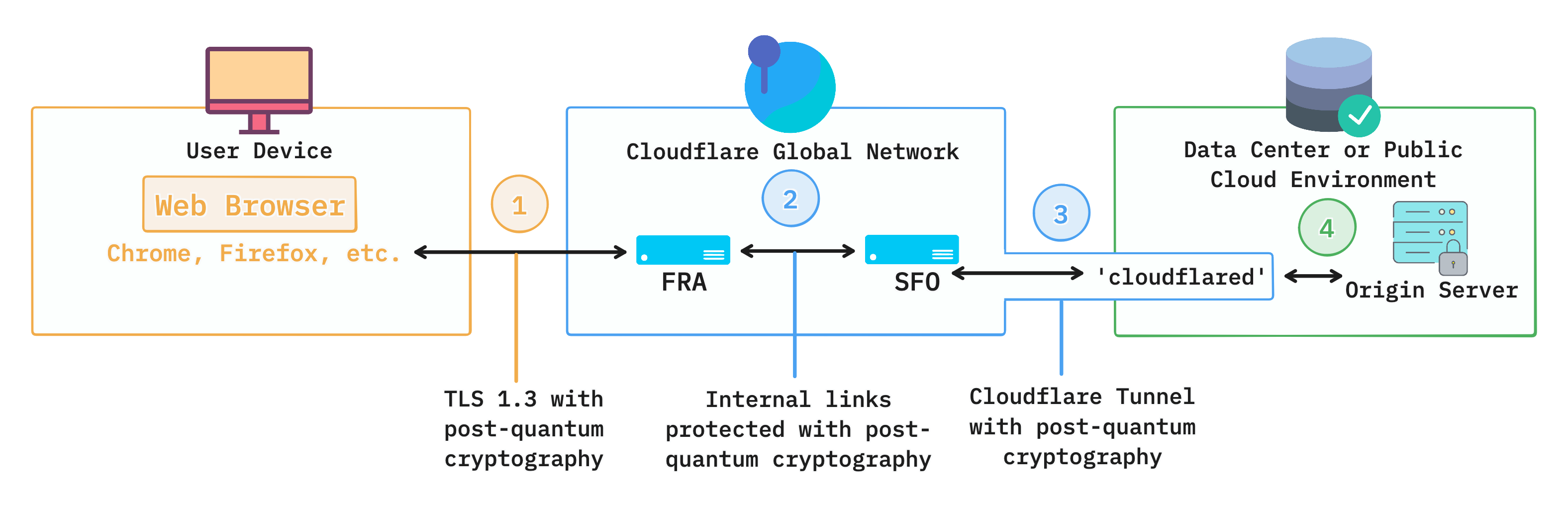}
    \caption{Cloudflare’s experimental deployment of PQC in the web ecosystem. 
The figure illustrates a user browser establishing a TLS~1.3 session with PQC-enabled key exchange (1), internal backbone links between Cloudflare data centers protected with PQC (2), Cloudflare Tunnels secured with PQC (3), and connections from edge servers to origin servers (4). 
These trials highlight the feasibility of hybrid PQC integration across end-user, backbone, and cloud environments while exposing practical challenges such as handshake size, latency, and certificate management \cite{Goldberg2025postquantum}.}
    \label{fig:cloudflare-pqc}
\end{figure}

The web ecosystem has been one of the earliest testing grounds for PQC adoption. Large-scale trials by Google and Cloudflare integrated CRYSTALS-Kyber into TLS handshakes (see~\autoref{fig:cloudflare-pqc}), demonstrating both the feasibility of PQC in production and the challenges posed by handshake size and latency overheads~\cite{cloudflare2024state,schwabe2021kemtls}. While these experiments confirmed that hybrid key exchange mechanisms can be deployed without breaking compatibility, they also revealed risks of inflated certificate chains and degraded user experience under high-latency conditions. The web domain illustrates the tension between forward security and real-time performance, making it an important driver of hybrid migration strategies.

\subsection{IoT and Embedded Systems}
IoT and embedded devices face perhaps the steepest PQC migration hurdles. These systems often operate under stringent memory, energy, and processing constraints, making large key sizes and computationally expensive operations difficult to accommodate~\cite{liu2024post,gsma2025PQIoT}. While hardware acceleration (via FPGA or ASIC co-design) has shown promise in reducing performance overhead, the lifecycle mismatch between PQC standards and long-lived embedded deployments remains problematic. Healthcare, IoT and critical infrastructure systems, such as industrial controllers, share these limitations: devices are frequently deployed for decades and cannot easily be retrofitted, raising the risk of widespread legacy exposure once quantum threats materialize. Consolidating IoT, healthcare, and critical infrastructure underlines the need for lightweight PQC profiles and domain-specific optimization.

\subsection{Financial Services and Central Bank Digital Currencies}

The financial sector faces significant systemic risks if PQC migration is poorly executed. Initiatives such as the Bank for International Settlements’ Project Leap highlight the importance of quantum-resilient infrastructures spanning the full financial ecosystem~\cite{auer2023projectLeap}. Central banks and regulators are increasingly aware that long-lived assets such as central bank digital currencies (CBDCs) are particularly exposed to quantum threats~\cite{nili2024cbdcQuantum}. Industry voices, including American Banker, emphasize that successful migration requires more than cryptographic substitution, demanding attention to compliance, operational costs, and international interoperability~\cite{pape2024beyondPQC}. Given the strict confidentiality requirements of financial records, early adoption of PQC in this sector is highly probable, though balancing efficiency with regulatory oversight remains a key challenge. 

\subsection{Blockchain and Decentralized Systems}

Decentralized networks encounter distinct challenges in transitioning to PQC, largely because consensus protocols and identity mechanisms depend heavily on digital signatures. Current schemes such as ECDSA and Edwards-Curve Digital Signature Algorithm (EdDSA) are quantum-vulnerable, and replacing them involves both technical migration and governance alignment across distributed communities~\cite{allende2023quantum}. Proposed pathways include deploying SPHINCS+ or lattice-based alternatives for wallet authentication and transaction signing, as well as hybrid solutions such as dual-signature blocks. However, upgrade cycles in blockchain environments are notoriously slow, leaving these ecosystems at heightened risk of “harvest now, decrypt later” attacks if quantum adversaries emerge before full PQC integration is achieved~\cite{olisa2025quantum}. 

\subsection{Cloud and Data Centers}

Major cloud providers such as Amazon Web Services (AWS), Azure, and Google are expected to play a leading role in early PQC adoption, given their central function in delivering cryptographic services a global scale. Deployment impacts span multiple layers of infrastructure, including TLS termination, data-at-rest encryption, key management, and isolation mechanisms for virtualized environments. The shared and multi-tenant design of cloud services magnifies performance concerns, since inefficient PQC algorithms could introduce latency across millions of users simultaneously. Ensuring crypto-agility within cloud application programming interfaces (APIs) and software development kits (SDKs) will therefore be essential to enable hybrid adoption and seamless protocol upgrades without service disruptions. 

\subsection{Regulatory and Policy Frameworks}

Policy directives and regulatory frameworks increasingly shape the timeline and scope of PQC deployment. In the United States, the NSA’s CNSA 2.0 establishes deadlines for federal migration~\cite{utimaco2024CNSA2}, while NIST’s NCCoE provides practical guidance on crypto-agility and phased adoption strategies~\cite{nist2023cryptoAgility}. In Europe, the Cyber Resilience Act and the European Union Agency for Cybersecurity (ENISA)’s recommendations add layers of compliance for operators of critical infrastructure. Sector-specific rules in domains such as healthcare and finance compound these requirements, raising the complexity of transition planning. At the international level, coordination remains limited, creating risks of fragmented timelines and interoperability barriers. 

\subsection{Synthesis and Lessons Learned}
Across domains, several cross-cutting themes emerge. First, lifecycle mismatch is a persistent issue: IoT, healthcare, and critical infrastructure devices may remain quantum-vulnerable well beyond PQC standardization. Second, governance bottlenecks, as seen in blockchain ecosystems, slow coordinated adoption even when technical solutions exist. Third, regulatory uncertainty and fragmented international timelines risk creating compliance burdens and interoperability failures. At the same time, sectors with concentrated infrastructure control-such as cloud providers and financial institutions, are positioned to lead PQC deployment. These contrasts suggest that PQC migration is not purely a technical challenge, but a socio-technical process shaped by policy, economics, and operational realities.

\section{Quantum Technologies for Security}
\label{sec:quantum-sec}

\subsection{Quantum Key Distribution (QKD)} 

QKD applies fundamental principles of quantum mechanics to achieve information-theoretic security. Unlike classical approaches that depend on computational hardness assumptions, QKD ensures that any eavesdropping attempt produces detectable disturbances, enabling both detection and mitigation \cite{mao2021recent}. The seminal BB84 protocol, introduced by Bennett and Brassard, laid the foundation for QKD by using photon polarization states to transmit key material \cite{mao2021recent, qian2025quantum}.

Modern QKD schemes fall into two main categories: discrete-variable QKD (DV-QKD), which relies on single-photon measurements, and continuous-variable QKD (CV-QKD), which uses quadrature measurements of light fields. DV-QKD offers strong theoretical guarantees, whereas CV-QKD is particularly attractive for deployment due to its compatibility with existing fiber-optic infrastructure and commodity optical devices such as lasers and detectors \cite{motaharifar2025survey}. Recent work has also integrated machine learning techniques into CV-QKD to enable adaptive noise suppression, parameter optimization, and automated system tuning \cite{motaharifar2025survey}.

\begin{figure}
    \centering
    \includegraphics[width=1\linewidth]{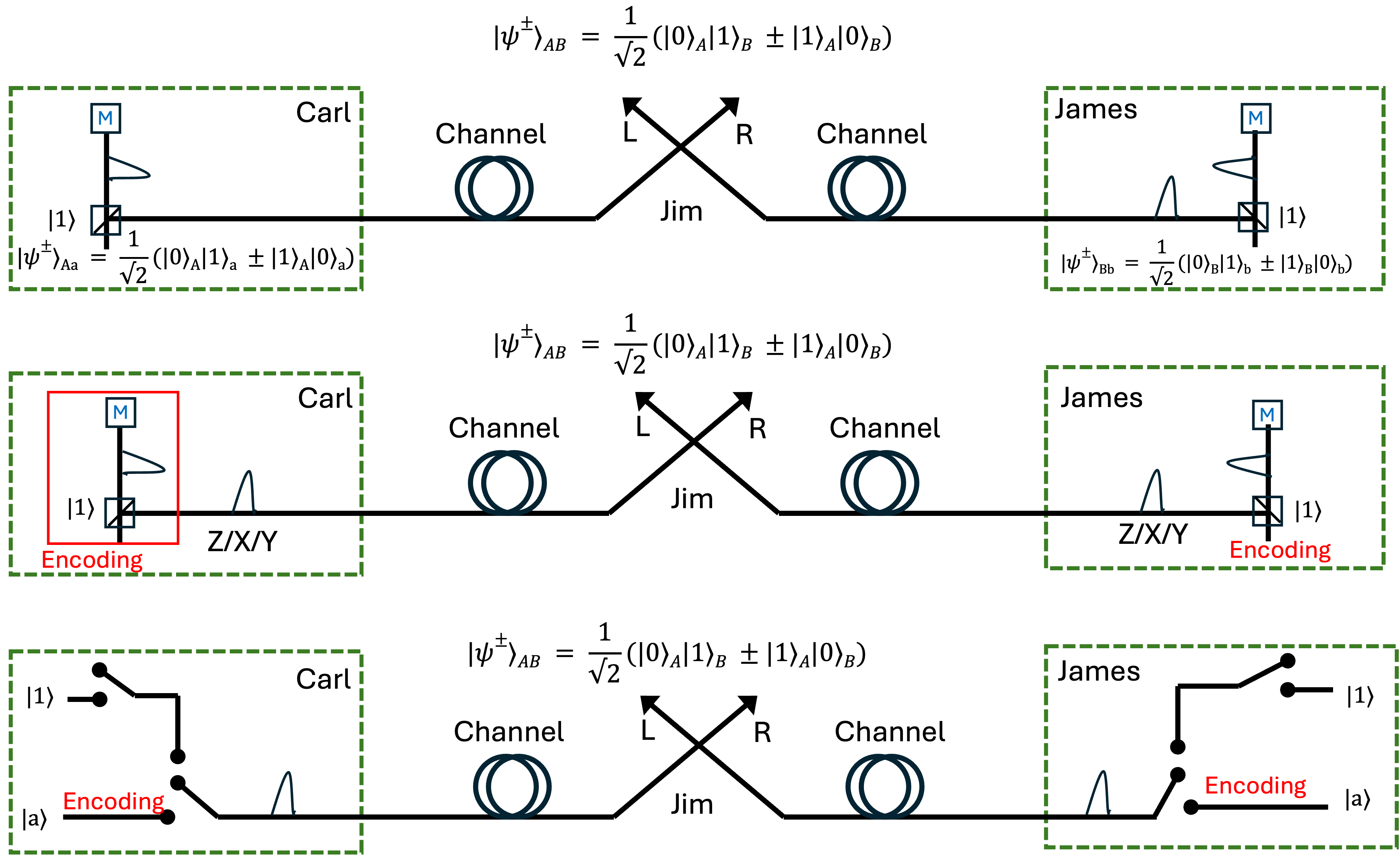}
    \caption{Illustration of twin-field QKD schemes that overcome the PLOB bound. 
    (a) Entanglement-based MDI-QKD with single-photon Bell state measurement (BSM): Carl and James each prepare a single-photon Bell state, while Charlie performs entanglement swapping. $M$ denotes the measurement basis (e.g., Z, X, or Y), which Carl and James apply after Charlie’s BSM. 
    (b) Prepare-and-measure MDI-QKD: Carl and James directly prepare qubits as superpositions of vacuum and one-photon states, performing $M$ before Charlie executes the single-photon BSM. 
    (c) Effective TF-QKD: indistinguishable photons from single-photon and laser sources enable long-distance interference. The single-photon source provides Z-basis encoding, while the laser source supports phase encodings (X and Y bases). Stable phase references are required for long-distance laser interference.}
    \label{fig:twin-file-qkd}
\end{figure}

Figure \ref{fig:twin-file-qkd} illustrates significant advances have extended the practical range of QKD. Twin-field QKD (TF-QKD) (see~\autoref{fig:twin-file-qkd}) overcame traditional distance limitations, enabling secure key distribution across more than 500 km of optical fiber, though with relatively low transmission rates \cite{mao2021recent}. Phase-matching QKD further improves efficiency and resilience, reinforcing its suitability for real-world communication networks \cite{mao2021recent}. 

Countermeasures against practical vulnerabilities have also matured. The decoy-state method defends against photon-number-splitting attacks by randomizing signal intensities \cite{sharma2021quantum}, while measurement-device-independent (MDI) QKD mitigates reliance on trusted detection devices, addressing a key class of side-channel threats \cite{mao2021recent}. Looking ahead, QKD protocols such as BB84 and E91 are expected to be integrated into emerging 6G infrastructures for securing financial transactions, government communications, and other critical systems \cite{qian2025quantum}. Nonetheless, long-distance QKD remains constrained by photon loss, motivating ongoing research into quantum repeaters and satellite-assisted channels as pathways to global-scale quantum-secure networks \cite{qian2025quantum}.

\subsection{QKD and PQC Complementarity} 
Although QKD and PQC pursue the same objective of quantum-safe security, they differ in scope and practicality. QKD provides unconditional confidentiality grounded in physical laws, but requires specialized hardware and is primarily limited to point-to-point links, resulting in high deployment costs and limited scalability \cite{garms2024experimental}. PQC, on the other hand, enables scalable authentication and confidentiality in large, heterogeneous networks using mathematically hard problems. However, PQC does not offer unconditional guarantees, and performance overheads may arise in latency-sensitive systems \cite{garms2024experimental, kavitha2025synergizing}.

By integrating both technologies, stronger layered security can be achieved. QKD can establish provably secure symmetric keys, while PQC provides scalable mechanisms for authentication and large-scale deployment \cite{hoque2024exploring, garms2024experimental}. Hybrid models ensure that as long as one of the building blocks-QKD, classical cryptography, or PQC-remains uncompromised, overall system security is preserved \cite{garms2024experimental}. This layered approach combines the high assurance of QKD with the scalability and interoperability of PQC, offering resilience against both current and future adversaries \cite{hoque2024exploring, kavitha2025synergizing}.

In practice, hybrid deployments of PQC and QKD can strengthen defenses against emerging cyber threats. While QKD ensures everlasting confidentiality in theory, PQC secures communication against realistic adversaries, including those wielding large-scale quantum computers. Together, they support the design of networks that are both theoretically robust and operationally feasible, enabling resilient infrastructures for government, finance, and critical services \cite{hoque2024exploring, rajkumar2024post, kavitha2025synergizing}.

\subsection{Quantum Random Number Generators (QRNGs)} 
High-quality randomness is essential for secure cryptographic primitives. Quantum Random Number Generators (QRNGs) exploit the inherent unpredictability of quantum processes, such as photon emission or phase fluctuations, to produce entropy that is provably resistant to prediction and replication \cite{pandey2024comparative, karera2024construction}. Unlike classical pseudo-random generators, QRNGs provide true randomness rooted in physical laws, ensuring robustness against adversaries with full knowledge of initial system states \cite{pandey2024comparative, karera2024construction}. Recent advances include semi-device-independent QRNGs (semi-DI QRNGs), which reduce reliance on device trust assumptions while maintaining verifiable security guarantees \cite{zhou2023numerical, lin2020security}. Similarly, source-independent QRNGs validate entropy independently of the quantum state source, thereby enhancing robustness against hardware imperfections \cite{zhou2024continuous, lin2020security}. Together, these paradigms improve the practicality of QRNG deployment across diverse environments.

Modern QRNGs have also achieved significant throughput improvements. FPGA-based and parallelized implementations now achieve rates exceeding 20 Gbps, enabling integration into real-time cryptographic systems and data centers \cite{guo2024parallel}. Advanced post-processing, including entropy extractors and error correction, ensures that generated sequences meet strict statistical requirements and resist residual correlations \cite{guo2024parallel}. Continuous-variable source-independent QRNGs further simplify system design while scaling to large problem sizes without reliance on specialized detectors \cite{zhou2024continuous}. Security remains contingent on mitigating implementation flaws. Imperfections in detectors, efficiency mismatches, and photon loss can degrade entropy quality and introduce exploitable patterns \cite{lin2020security}. To safeguard against such risks, QRNG outputs are routinely tested with standard batteries (e.g., NIST, Diehard, ENT Randomness Test Suite (ENT)) and supplemented with error correction and extraction techniques \cite{pandey2024comparative, karera2024construction}. These advances position QRNGs as the state-of-the-art solution for applications requiring provably secure randomness, from cryptography to large-scale simulations \cite{guo2024parallel, karera2024construction}.

\subsection{Limitations and Reality Check} 
Despite rapid progress, quantum technologies face significant barriers to widespread adoption. Technical, economic, and implementation-related challenges limit their near-term practicality.

\subsubsection{Hardware and Scalability Constraints}
Current quantum computers, commonly referred to as noisy intermediate-scale quantum (NISQ) devices, suffer from limited qubit counts, susceptibility to noise, and error-prone operation, which restricts their ability to perform large-scale tasks reliably \cite{kundu2024sok}. For example, even systems with 100 qubits cannot effectively execute algorithms of equivalent scale due to error accumulation, forcing researchers to downscale problems substantially \cite{kundu2024sok, akter2023exploring}.

\subsubsection{Economic and Temporal Barriers}
Quantum infrastructure is also cost-prohibitive. Commercial services, such as IBM’s quantum cloud, are priced at approximately \$1.60 per second, which is over 2,300 times more expensive than comparable classical GPU resources \cite{kundu2024sok}. Moreover, limited access to hardware leads to long job queues, with training of quantum machine learning models potentially requiring months of runtime \cite{kundu2024sok}. These costs and delays make large-scale adoption impractical in cost-sensitive applications.

\subsubsection{Implementation Gaps and Device Imperfections}
Quantum-enhanced systems, despite their theoretical robustness, remain susceptible to a range of practical vulnerabilities. Both classical and quantum machine learning models can be compromised by adversarial perturbations that degrade performance and reliability in security-critical contexts \cite{akter2023exploring}. Cloud-based quantum services add further exposure by introducing risks of data leakage and potential manipulation of quantum circuits. At the hardware level, crosstalk between qubits and other device imperfections can create avenues for fault injection attacks, threatening the integrity of computations and overall system reliability \cite{kundu2024sok}. Similarly, imperfections in measurement devices directly affect the security of quantum random number generators (QRNGs). Detector noise, efficiency mismatches, and photon leakage can reduce entropy and introduce statistical biases \cite{cao2024deep}. These flaws underscore the need for rigorous calibration, continuous monitoring, and robust post-processing to maintain the quality of random outputs. Without such safeguards, QRNGs risk generating predictable sequences that undermine the cryptographic strength of dependent systems \cite{cao2024deep}.

\section{Comparative Analysis and Discussion}
\label{sec:comp-analysis}

The comparative evaluation 
of PQC families highlights their relative strengths, limitations, and deployment readiness. This discussion synthesizes mathematical foundations, cryptanalytic track records, performance characteristics, and standardization progress through 2025. The goal is to provide a balanced perspective that guides both near-term adoption and long-term resilience planning. The comparison in~\autoref{tab:comp-ana} consolidates the defining traits of each post-quantum cryptographic family. Lattice-based constructions emerge as the most mature, balancing efficiency, security, and implementability, and forming the foundation of the current NIST standards. Code-based schemes remain time-tested but face practical constraints from key size and limited applicability. Hash-based designs, while conservative, provide unmatched confidence in long-term security and are ideal for constrained verification scenarios. Multivariate and MPC-in-the-Head approaches illustrate active research directions focused on reducing footprint and improving fault tolerance, whereas isogeny-based systems, once considered promising, have been largely deprecated following recent cryptanalytic breaks. Overall, this comparative synthesis underscores that PQC evolution is characterized by a trade-off between mathematical diversity, operational efficiency, and long-term confidence, with lattice-based and hash-based schemes currently leading the path toward wide-scale deployment.

\begin{table*}[ht]
\small
\centering
\caption{Comprehensive PQC Category Comparison}
\label{tab:comp-ana}
\renewcommand{\arraystretch}{1.2}
\setlength{\tabcolsep}{3pt}
\begin{tabular}{P{2cm}P{3cm}P{4cm}P{4cm}}
\toprule
\rowcolor{tabheader}%
    \textcolor{white}{\textbf{Category}} &
    \textcolor{white}{\textbf{Core Idea}} &
    \textcolor{white}{\textbf{Pros / Cons}} &
    \textcolor{white}{\textbf{Status}} 
    \\
\midrule
Lattice-Based \newline \cite{regev2009lattices, langlois2015worst, cherkaoui2024exploring} 
& LWE, M-LWE, NTRU; ML-KEM, ML-DSA, FALCON 
& Pros: Strong worst-case hardness; efficient and versatile implementations. \newline Cons: Medium key sizes; side-channel vulnerabilities. 
& 20+ years analysis; FIPS 203/204/206; Production ready \\
\midrule
Code-Based \newline \cite{melchor2018hamming, nguyen2019new} 
& Syndrome decoding; HQC, McEliece, BIKE 
& Pros: Long security record; independent assumptions. \newline Cons: Very large keys; limited signature options. 
& 45+ years secure; HQC finalist; Limited deployment \\
\midrule
Hash-Based \newline \cite{nguyen2019new} 
& Hash preimage resistance; SPHINCS+, eXtended Merkle Signature Scheme (XMSS) 
& Pros: Minimal assumptions; strong long-term confidence. \newline Cons: Large signatures (8–30 KB); slower signing. 
& Conservative; FIPS 205; Specialized use \\
\midrule
Multivariate \newline \cite{beullens2022breaking, kramer2019fault, dey2023progress} 
& MQ equations; UOV, MAYO, Rainbow (broken) 
& Pros: Compact signatures; efficient verification. \newline Cons: Vulnerable to advanced algebraic attacks. 
& Compromised; active research; Evaluation phase \\
\midrule
Isogeny-Based \newline \cite{de2022zero, weitkamper2023cryptanalysis} 
& Supersingular isogenies; SIKE (broken), Commutative Supersingular Isogeny Diffie–Hellman (CSIDH) 
& Pros: Extremely small key sizes (pre-break). \newline Cons: Cryptanalytically broken in 2022. 
& Eliminated from NIST process; Not viable \\
\midrule
MPC-in-Head \newline \cite{nguyen2019new} 
& ZK proofs, symmetric primitives; Picnic, SDITH 
& Pros: Flexible assumptions; avoids algebraic structures. \newline Cons: Very large signatures; high memory demand. 
& Early stage; research only; Experimental \\
\bottomrule
\end{tabular}

\vspace{2mm}
\begin{minipage}{\textwidth}
\footnotesize \textit{Note:} Summary reflects research through 2025 and NIST PQC standardization progress.
\end{minipage}
\end{table*}

\subsection{Security Analysis and Cryptanalytic Evolution}

Lattice-based cryptography has emerged as the most mature and reliable family, combining rigorous worst-case hardness guarantees with practical efficiency. These schemes have undergone over two decades of intensive scrutiny, and the standardization of ML-KEM, ML-DSA, and FN-DSA as FIPS reflects their readiness for widespread deployment \cite{regev2009lattices}. Their adoption provides production-ready solutions that balance theoretical soundness and practical performance.

Code-based schemes contribute indispensable algorithmic diversity by relying on an entirely independent mathematical foundation. HQC’s progress toward standardization in 2025 positions it as a critical fallback to lattice-based approaches \cite{melchor2018hamming}. The enduring security of Classic McEliece underscores the conservative reliability of this family, despite the cost of very large public keys.

Hash-based signatures provide the most conservative foundation, with security depending solely on the properties of well-analyzed cryptographic hash functions \cite{nguyen2019new}. While performance penalties from large signature sizes and slower signing limit their use in interactive protocols, they remain essential for applications requiring long-term assurance against unforeseen advances in algebraic cryptanalysis.

In contrast, multivariate and isogeny-based families have suffered significant setbacks. The rectangular MinRank attack on Rainbow demonstrated the susceptibility of multivariate schemes to advanced algebraic methods \cite{beullens2022breaking}, while the collapse of SIKE illustrated how subtle structural vulnerabilities can be exploited using unexpected mathematical insights \cite{de2022zero}. These failures highlight the need for sustained scrutiny before non-lattice families can be considered for critical deployments.

\subsection{Deployment Recommendations and Strategic Considerations}

The comparative analysis suggests that lattice-based schemes should serve as the primary foundation for post-quantum migration. ML-KEM provides the optimal choice for key encapsulation due to its strong security guarantees and efficient performance, while ML-DSA offers balanced signature sizes and signing efficiency suitable for general-purpose use. FN-DSA, with its compact signatures, is well-suited for bandwidth-constrained environments, though its implementation complexity and side-channel sensitivity require careful handling. SLH-DSA, despite slower performance and larger signatures, remains indispensable as a conservative option for high-assurance domains.

Strategic deployment requires not only algorithm selection but also infrastructure planning. Organizations should adopt crypto-agility frameworks that support seamless algorithm transitions as new cryptanalytic results emerge. Migration strategies must account for performance trade-offs, implementation security (particularly side-channel resistance), and the operational impact of larger keys and signatures. In practice, this means prioritizing lattice-based standards for near-term adoption, while preparing code-based and hash-based mechanisms as resilient alternatives. Multivariate and isogeny-based designs, though currently compromised, still contribute to the diversity of research and may inform future paradigms. By combining standardized algorithms with robust agility frameworks, organizations can ensure both immediate protection and adaptability against evolving quantum-era threats.

In summary, the comparative analysis demonstrates that lattice-based schemes should serve as the cornerstone of post-quantum deployment, with code-based and hash-based schemes providing essential algorithmic diversity for resilience against future advances in cryptanalysis. While multivariate and isogeny-based families have been weakened by recent breakthroughs, they underscore the importance of continuous evaluation and the need for flexible, agile frameworks that can accommodate both emerging threats and novel algorithmic designs. This strategic balance between immediate deployment readiness and long-term adaptability directly motivates the discussion of open problems in the next section.

\section{Open Problems and Future Research}
\label{sec:open-problems}

While the standardization of CRYSTALS-Kyber, CRYSTALS-Dilithium, and SPHINCS+ represents a major milestone for PQC, significant challenges remain before the ecosystem can be considered secure, efficient, and universally deployable.

First, questions persist regarding the long-term cryptanalytic maturity of candidate algorithms. Lattice-based schemes currently dominate, but their resilience depends on assumptions that could be weakened by advances in lattice reduction or quantum algorithms~\cite{regev2009lattices, langlois2015worst}. Other families, including multivariate and isogeny-based cryptography, have already experienced major setbacks after high-profile breaks~\cite{beullens2022breaking, castryck2023efficient, skuggedal_bsidh_attack_2023}, underscoring the importance of deeper theoretical analysis and diversification beyond a narrow set of assumptions. Second, side-channel resistance and implementation security remain critical gaps. Even standardized schemes such as Dilithium have been shown to be vulnerable to sophisticated side-channel and fault-injection attacks under improper deployment conditions~\cite{zeitoun2022sidechannel, liu2025release}. The design of lightweight, provably secure countermeasures that preserve efficiency is particularly pressing for hardware accelerators and embedded systems. Third, the challenge of crypto-agility and migration pathways is unresolved. As emphasized by NIST and industry groups~\cite{nist2023cryptoAgility, zacharopoulos2024drawbacks}, replacing classical algorithms in large-scale infrastructures such as TLS, PKI, and financial systems will require phased deployment strategies, hybrid solutions, and continuous adaptability. Research into frameworks that enable seamless algorithm transitions while minimizing risks of misconfiguration, certificate bloat, and downgrade attacks remains a priority.

Fourth, domain-specific adoption barriers demand targeted solutions. IoT and embedded platforms face severe resource constraints that complicate PQC integration without specialized optimizations~\cite{liu2024post, gsma2025PQIoT}. Financial systems and CBDCs must address systemic risks and regulatory uncertainties~\cite{auer2023projectLeap, nili2024cbdcQuantum}, while blockchain ecosystems confront governance and consensus challenges when integrating new primitives~\cite{allende2023quantum, olisa2025quantum}. Each sector will require customized performance benchmarks, compliance frameworks, and hardware–software co-design methodologies. Fifth, the relationship between PQC and QKD remains underexplored. Although QKD offers information-theoretic security, deployment is constrained by infrastructure cost, distance limitations, and interoperability challenges~\cite{garms2024experimental, garg2024post}. Future research should investigate hybrid trust models that combine the scalability of PQC with the redundancy and high assurance of QKD, particularly in critical infrastructure domains.

Finally, there is a growing need for comprehensive benchmarking and real-world validation. Existing performance studies~\cite{demir2025performance, commey2025performance} remain confined to controlled laboratory settings. Large-scale, heterogeneous testbeds spanning data centers, mobile networks, IoT deployments, and financial platforms are essential to evaluate PQC performance under operational constraints and to inform practical optimization strategies. Overall, PQC research is shifting from an algorithm-centric paradigm toward a broader systems and deployment perspective. Progress will require simultaneously reinforcing the theoretical foundations of candidate families, strengthening implementation security, and developing migration strategies that accommodate the diversity of real-world infrastructures. Addressing these open problems will shape the trajectory of PQC in the coming decade and determine its readiness for global adoption.

\section{Conclusions}
\label{sec:conclusion}

PQC has moved from a niche academic pursuit to a pressing global security priority, driven by the rapid progress of quantum computing and the vulnerabilities of widely deployed public-key systems such as RSA and ECC. This survey has contributed by systematically classifying PQC algorithmic families, analyzing their strengths, weaknesses, and trajectories through the NIST standardization process, and examining the domain-specific challenges in real-world deployment. We have also highlighted the role of hybrid approaches and crypto-agility frameworks as transitional strategies, ensuring that organizations can prepare for quantum threats without sacrificing interoperability or performance in the near term. Collectively, this work provides an integrated view of PQC as both a technical and socio-technical phenomenon, bridging algorithmic innovation and system-level implementation.

Despite these contributions, several limitations of the current body of research must be acknowledged. First, while the NIST process has delivered three standardized algorithms (Kyber, Dilithium, and SPHINCS+), the comparative evaluation of alternative schemes-particularly multivariate, code-based, and isogeny-based approaches-remains incomplete. Many existing studies rely on benchmarks under specific conditions that may not generalize to diverse environments such as embedded IoT devices, cloud-scale infrastructures, or mission-critical systems with strict latency requirements. Second, side-channel resistance and implementation security are often treated as afterthoughts in algorithmic proposals, leaving open questions about their viability in adversarial real-world contexts. Third, the literature surveyed reveals a geographic and institutional concentration of research, which may limit the diversity of perspectives on deployment models and governance frameworks. Acknowledging these limitations clarifies the scope of the current survey and signals the need for broader and deeper engagement across the PQC research ecosystem.

Future work in PQC must therefore pursue three parallel tracks. On the algorithmic side, research should continue to explore diversity beyond the lattice- and hash-based paradigms that dominate current standards. Code-based and multivariate cryptography, though less mature, offer valuable redundancy in the event of unforeseen cryptanalytic breakthroughs. Establishing rigorous security proofs under realistic quantum adversary models remains a critical agenda item, as do efforts to refine parameter sets to balance efficiency with conservative security margins. On the system side, performance optimization, integration in constrained environments, and resilience against side-channel attacks are urgent areas of study. The need for efficient hardware accelerators, lightweight implementations, and verified libraries will only grow as PQC moves from pilot deployments to mainstream adoption. Finally, on the policy and governance side, the global synchronization of migration strategies is essential to prevent fragmentation, especially in sectors like finance, telecommunications, and defense, where interoperability is paramount.

Another crucial area for future inquiry is the interaction between PQC and complementary quantum-safe technologies. QKD, though not a substitute for PQC, may play a complementary role in specific high-assurance contexts. Similarly, blockchain and distributed ledger systems provide fertile ground for PQC research, as their long-lived security assumptions and decentralized architectures demand both forward secrecy and scalability. The integration of PQC into emerging paradigms such as zero-trust architectures, homomorphic encryption, and secure multiparty computation also warrants systematic investigation, as these fields converge in building the security foundations of a post-quantum digital ecosystem.
Finally, researchers must recognize that PQC migration is not solely a technical problem but a socio-technical transition akin to the historical adoption of public-key cryptography in the late 20th century. Success will require coordination between researchers, standards bodies, policymakers, and industry stakeholders. Challenges such as “harvest now, decrypt later” threats, uneven adoption across regions, and the economic costs of retrofitting infrastructure highlight the importance of developing migration playbooks, sector-specific roadmaps, and global collaboration mechanisms. By foregrounding these systemic issues, future work can ensure that PQC research translates into secure, deployable, and equitable infrastructures worldwide.

In conclusion, this survey provides a structured foundation for understanding the state of PQC at a pivotal historical moment. Its contributions lie in synthesizing algorithmic advances, contextualizing system-level challenges, and identifying open research problems. Its limitations underscore the need for continued, diverse, and globally coordinated inquiry. The future of PQC will be shaped not only by cryptographic breakthroughs but also by engineering ingenuity, cross-sectoral collaboration, and sustained preparedness. If these elements align, the transition to quantum-safe security will be achievable, ensuring that the digital systems underpinning modern society remain resilient in the quantum era. Post-quantum cryptography has transitioned from theoretical promise to standardized reality. Sustained collaboration among academia, industry, and government is now essential to ensure secure, efficient, and verifiable migration toward quantum-safe digital infrastructures.

\bibliographystyle{ACM-Reference-Format}
\bibliography{pqc-checked}

\appendix

\section{List of Acronyms}

To assist the reader, a list of all acronyms and abbreviations used throughout this paper is provided in Table \ref{tab:acronym}. This includes terms related to current cryptographic standards, PQC candidates, and quantum-safe communication technologies.

\begin{table}[ht]
\centering
\small
\caption{List of Acronyms Used in This Paper}
\label{tab:acronym}
\begin{tabular}{ll}
\toprule
\rowcolor{tabheader}%
    \textcolor{white}{\textbf{Acronym}} &
    \textcolor{white}{\textbf{Full Form}} \\
\midrule
CBDC   & Central Bank Digital Currency \\
CNSA 2.0 & Commercial National Security Algorithm Suite 2.0 \\
CRYSTALS-Dilithium & Cryptographic Suite for Algebraic Lattices – Dilithium (digital signature) \\
CRYSTALS-Kyber & Cryptographic Suite for Algebraic Lattices – Kyber (key encapsulation) \\
CV-QKD & Continuous-Variable Quantum Key Distribution \\
DNSSEC & Domain Name System Security Extensions \\
DV-QKD & Discrete-Variable Quantum Key Distribution \\
ECDSA  & Elliptic Curve Digital Signature Algorithm \\
FALCON & Fast Fourier Lattice-based Compact Signatures over NTRU \\
FFT    & Fast Fourier Transform \\
FIPS   & Federal Information Processing Standard \\
FORS   & Forest of Random Subsets (component in SPHINCS+) \\
FPGA   & Field-Programmable Gate Array \\
FN-DSA & Falcon Digital Signature Algorithm (NTRU lattice-based) \\
HQC    & Hamming Quasi-Cyclic (code-based KEM) \\
HSM    & Hardware Security Module \\
IoT    & Internet of Things \\
KEMTLS & Key Encapsulation Mechanism-based TLS \\
ML-KEM & Module-Lattice Key Encapsulation Mechanism (CRYSTALS-Kyber) \\
ML-DSA & Module-Lattice Digital Signature Algorithm (CRYSTALS-Dilithium) \\
MLWE   & Module Learning With Errors \\
MiRitH & MPC-in-the-Head with Repeated Iterations of Threshold Hashing \\
MPC    & Multi-Party Computation \\
MPC-in-the-Head & Zero-knowledge based PQ signature construction \\
NCCoE  & National Cybersecurity Center of Excellence \\
NIST   & National Institute of Standards and Technology \\
PERK   & Practical Efficient Randomized MPC-in-the-Head with K-projections \\
PKI    & Public Key Infrastructure \\
PQC    & Post-Quantum Cryptography \\
QKD    & Quantum Key Distribution \\
QRNG   & Quantum Random Number Generator \\
RSA    & Rivest–Shamir–Adleman (public-key cryptosystem) \\
SHA    & Secure Hash Algorithm \\
SHAKE  & Secure Hash Algorithm Keccak (extendable-output functions) \\
SIDH   & Supersingular Isogeny Diffie–Hellman \\
SIKE   & Supersingular Isogeny Key Encapsulation (broken scheme) \\
SLH-DSA & Stateless Hash-Based Digital Signature Algorithm (SPHINCS+) \\
TLS    & Transport Layer Security \\
UOV    & Unbalanced Oil and Vinegar (multivariate signature scheme) \\
UDP    & User Datagram Protocol \\
\bottomrule
\end{tabular}
\end{table}

\end{document}